\documentclass[preprint2,letterpaper]{emulateapj}

\slugcomment{}

\shorttitle{X-ray spectra of the most X-ray luminous radio-quiet RBS-QSOs}
\shortauthors{Krumpe et al.}

\begin{document}

%%%%%%%%%%%%%%%%%%%%%%%%%%%%%%%%%%%%%%%%%%%%%%%%%%%%%%%%%%%%%%%%%%%%%%%%%%%%%%%%%%%%%%%%%%%%%%%%%%%%%%%%%%%%%%%%%%%%%%%%%%%%%%%%%%%%%%%%%%
\title{The {\em XMM-Newton} X-ray Spectra of the Most X-ray Luminous Radio-quiet
{\em ROSAT} Bright
Survey-QSO{\scriptsize s}: A Reference Sample for the Interpretation of
High-redshift QSO Spectra}

\author{M. Krumpe\altaffilmark{1}, G. Lamer\altaffilmark{2}, A.
Markowitz\altaffilmark{1}, and A. Corral\altaffilmark{3}}

\altaffiltext{1}{University of California, San Diego, Center for Astrophysics and
                Space Sciences, 9500 Gilman Drive, La Jolla, CA 92093-0424, USA}
\altaffiltext{2}{Astrophysikalisches Institut Potsdam, An der Sternwarte 16,
                 14482 Potsdam, Germany}
\altaffiltext{3}{INAF - Osservatorio Astronomico di Brera, via Brera 28, 20121
Milan, Italy}

\email{mkrumpe@ucsd.edu}

%%%%%%%%%%%%%%%%%%%%%%%%%%%%%%%%%%%%%%%%%%%%%%%%%%%%%%%%%%%%%%%%%%%%%%%%%%%%%%%%%%%%%%%%%%%%%%%%%%%%%%%%%%%%%%%%%%%%%%%%%%%%%%%%%%%%%%%%%%
\begin{abstract}

We present the broadband X-ray properties of four of the most X-ray luminous
($L_{\rm X} \ge 10^{45}$ erg s$^{-1}$ in the 0.5--2 keV band) radio-quiet QSOs 
found in the {\em ROSAT} Bright Survey. This uniform sample class, which 
explores the extreme end of the QSO luminosity function, exhibits surprisingly 
homogenous X-ray spectral properties: a soft excess with an extremely smooth 
shape containing no obvious discrete features, a hard power law above 2 keV,  
and a weak narrow/barely resolved Fe K$\alpha$ fluorescence line for the three high 
signal-to-noise ratio (S/N) spectra. The soft excess can be well fitted 
with only a soft power law. No signatures of warm or cold intrinsic absorbers are 
found. 

The Fe K$\alpha$ centroids and the line widths indicate emission from
neutral Fe ($E=6.4$ keV) originating from cold material from distances of only a few 
light days or further out. The well-constrained equivalent 
widths (EW) of the neutral Fe lines are higher than expected from the X-ray Baldwin 
effect which has been only poorly constrained at very high luminosities. Taking into 
account our individual EW measurements, we show that the X-ray Baldwin effect flattens
above $L_{\rm X} \sim 10^{44}$ erg s$^{-1}$ (2--10 keV band) where an almost constant
$\langle$EW$\rangle$ of $\sim$100 eV is found.

We confirm the assumption of having very similar X-ray AGN properties when interpreting 
stacked X-ray spectra. Our stacked spectrum serves as a superb reference for the interpretation of low S/N 
spectra of radio-quiet QSOs with similar luminosities at higher redshifts routinely 
detected by {\em XMM-Newton} and {\em Chandra} surveys.

\end{abstract}

%%%%%%%%%%%%%%%%%%%%%%%%%%%%%%%%%%%%%%%%%%%%%%%%%%%%%%%%%%%%%%%%%%%%%%%%%%%%%%%%%%%%%%%%%%%%%%%%%%%%%%%%%%%%%%%%%%%%%%%%%%%%%%%%%%%%%%%%%%

%%%%%%%%%%%%%%%%%%%%%%%%%%%%%%%%%%%%%%%%%%%%%%%%%%%%%%%%%%%%%%%%%%%%%%%%%%%%%%%%%%%%%%%%%%%%%%%%%%%%%%%%%%%%%%%%%%%%%%%%%%%%%%%%%%%%%%%%%%
\section{Introduction}

Active galactic nuclei (AGNs) are classified by luminosity and by
phenomenological criteria. The most luminous objects are the
radio-loud quasars (quasi-stellar radio sources) and the radio-quiet
QSOs (quasi-stellar objects). Radio-loud objects are characterized
by relativistic jets. Since the jet emission of the radio-loud objects
is likely to dilute the X-ray spectrum from the central engine of the
AGN, radio-quiet objects are more suitable for the study of X-ray emission
associated with the accretion disk.

To date, detailed information on the X-ray spectra of radio-quiet AGNs
has largely been derived from observations of nearby flux-bright objects.
Spectroscopy of low luminosity ($L_{\rm X} < 10^{44}$\,erg s$^{-1}$) radio-quiet
Seyfert galaxies has revealed a relatively complex multi-component X-ray 
spectrum. In the soft X-rays, a steep component (soft excess) has been observed in many
cases. In the past this component has been interpreted as the tail of the thermal
emission from the accretion disk or Comptonization of EUV accretion
disk photons (\citealt{arnaud_raymont_1985};
\citealt{kawaguchi_shimura_2001}; \citealt{porquet_reeves_2004}).
On the other hand, \cite{crummy_fabian_2006} and \cite{ross_fabian_2005} discuss
ionized disk reflection as an explanation. The spectrum
at soft X-rays can also be influenced by a cold or warm absorber.
Following this observation, \cite{gierlinski_done_2004} consider absorption
along the line of sight from relativistically outflowing, optically thin, warm gas
as an explanation for the soft excess. At higher energies, X-ray spectra are
dominated by a hard ($\Gamma$$\sim$2) power law, attributed to inverse Compton
emission from a hot, thin corona surrounding the accretion disk.
Furthermore, Compton reflection of the primary flux by the disk or torus
is inferred by the detection of a 6.4 keV
fluorescent Fe K$\alpha$ line and the 7.1 keV Fe absorption edge in
AGN spectra. The first narrow, unresolved Fe K$\alpha$ line
in an AGN was observed in {\em OSO-8} spectra of Centaurus A by
\cite{mushotzky_serlemitsos_1978}. Various {\em XMM-Newton} and {\em Chandra}
observations have shown that a narrow
(FWHM up to $\sim$ $10^{3}$ km s$^{-1}$ due to Doppler broadening) Fe line
component is virtually ubiquitous in Seyfert spectra and indicates emission from
gas roughly commensurate with the optical Broad Line Region (e.g.,
\citealt{nandra_2006}; \citealt{shu_yaqoob_2010}) and/or emission from the
putative Compton-thick, homogeneous molecular
torus invoked in Seyfert type I/II Unification schemes and
thought to be located about light-years from the supermassive black hole
(e.g., \citealt{antonucci_1993}).
A relativistically broadened (FWHM$ \sim 10^5$ km s$^{-1}$) component
indicates emission from within a few $R_{\rm Sch}$ ($\equiv 2GM_{\rm
BH}/c^2$) of the black hole where $M_{\rm BH}$ is the black hole mass (e.g.,
\citealt{tanaka_nandra_1995};
\citealt{fabian_vaughan_2002}) but appears in only about half
of Seyferts (\citealt{nandra_oneill_2007}).

There is evidence for the narrow Fe K$\alpha$ equivalent width to be
anti-correlated with X-ray luminosity, the so-called X-ray Baldwin effect.
This effect has been documented in several AGN samples, including
\cite{iwasawa_taniguchi_1993} using {\em Ginga}, \cite{nandra_george_1997}
using {\em ASCA}, and \cite{page_obrien_2004a, page_reeves_2005} using
{\em XMM-Newton}. However, the fitted correlation is mainly due to
$L_{\rm 2-10} < 10^{45}$\,erg s$^{-1}$ objects (\citealt{bianchi_guainazzi_2007}).

Improving the signal-to-noise ratio (S/N) by stacking the X-ray spectra of
high-redshift AGNs is a common method for high-redshift AGNs.
\cite{streblyanska_hasinger_2005} stack the {\em XMM-Newton}
spectra of AGNs in the Lockman Hole survey, which covers a large range of
luminosities and redshifts. In the average spectra of both
type I and type II objects they find significant Fe fluorescent emission,
which seemed to be relativistically blurred (type I AGNs: EW$\sim$600 eV,
type II AGNs: EW$\sim$400 eV). In comparison, recent studies such as
the {\em XMM-Newton} Wide Angle Survey by \cite{corral_page_2008} find
only a narrow unresolved Fe line of EW$\sim$90 eV in the stacked spectrum
of 606 type I AGNs. \cite{mateos_carrera_2010} determine the mean
continuum shape of 487 AGNs in the {\em XMM-Newton} Wide Angle Survey by
stacking the spectra at different cosmic epochs.

The {\em ROSAT} Bright Survey (RBS; \citealt{schwope_hasinger_2000})
identifies the brightest (0.1--2.4 keV count rate $>$0.2\,ct s$^{-1}$,
flux limit $f_{\rm 0.5-2} \sim 10^{-12}$ erg cm$^{-2}$ s$^{-1}$)
X-ray sources from the {\em ROSAT} All-Sky Survey (RASS) at high Galactic latitudes.
X-ray luminous radio-quiet QSOs are rare locally and their spectral properties are
so far
poorly established. Due to their high X-ray flux, the RBS objects form an ideal sample
to obtain the most statistically significant X-ray spectra and allows us to evaluate
directly their X-ray spectral properties and the properties ($\sigma$, EW) of their
Fe K$\alpha$ lines. In this paper, we use {\em XMM-Newton} to
study the spectral properties of four objects that belong to the
12 most X-ray luminous radio-quiet QSOs found in the RBS with
$L_{\rm 0.5-2} \ge 10^{45}$ erg s$^{-1}$. Such analysis of the most luminous
RBS-QSOs
will test the assumption of having very similar X-ray AGN properties when
stacking their X-ray spectra. The sample will serve as a medium-$z$ reference for the
interpretation of low S/N QSO spectra at high redshifts.
In addition, the EW of the Fe K$\alpha$ line will be directly accessible in the
individual
X-ray spectra. This will add several data points in the poorly studied high
luminosity range
of the X-ray Baldwin effect.

The paper is organized as follows. In Section~2 we describe how we selected the sample.
Section~3 gives black hole mass and Eddington ratio estimates from optical spectra.
In Section~4 the X-ray data reduction is specified, and we present the analysis of the
X-ray data for RBS 1055, RBS 320, RBS 1124, and RBS 1883 in Section~5. In Section~6 we derive 
the optical-to-X-ray spectral indices for our objects.
Section~7 comments on the long-term flux variability of the RBS-QSOs.
The construction and analysis of the averaged spectrum of the most X-ray luminous 
radio-quiet RBS-QSOs is given in Section~8. 
In Section~9 we discuss the results in the context of other studies,
and we give our conclusions in Section~10.

All quoted uncertainties represent 90\% confidence intervals
($\Delta \chi^2=2.706$ for one parameter) unless otherwise stated. 
We adopt a standard flat $\Lambda$$CDM$ cosmology with
$\Omega_{\rm M}$ = 0.3, $\Omega_{\Lambda}$ =0.7, $H_0$ = 70\,km\,s$^{-1}$\,Mpc$^{-1}$
(\citealt{spergel_verde_2003}).

%%%%%%%%%%%%%%%%%%%%%%%%%%%%%%%%%%%%%%%%%%%%%%%%%%%%%%%%%%%%%%%%%%%%%%%%%%%%%%%%%%%%%%%%%%%%%%%%%%%%%%%%%%%%%%%%%%%%%%%%%%%%%%%%%%%%%%%%%%
\vspace*{0.1in}
\section{Construction a Sample of the Most X-ray Luminous Radio-quiet RBS-QSO{\scriptsize s}}

Based on the RASS (\citealt{voges_aschenbach_1999}),
\cite{schwope_hasinger_2000} present a spectroscopic identification
catalog of high-Galactic ($\mid$$b$$\mid > 30^{\circ}$) latitude RASS
sources with 0.1--2.4 keV count rates above 0.2 s$^{-1}$. The identification
program of 2042 RBS sources reaches a spectroscopic completeness ratio of
more than 99.5\%.

We preselect objects that are classified as AGNs and sort them in descending
order of their Galactic-absorption corrected 0.5--2 keV luminosity.
BL Lac objects are not considered for this
study. Since we are interested only in radio-quiet AGNs, we furthermore exclude
radio-loud AGNs by their radio-to-optical flux density $R>10$
(\citealt{kellermann_sramek_1989}). The NASA/IPAC Extragalactic
Database\footnote{\tt http://nedwww.ipac.caltech.edu} (NED) is
used to obtain radio detections and fluxes. In addition, we
check the object's coordinates for entries in the radio catalogs
NRAO VLA Sky Survey (NVSS; \citealt{condon_cotton_1998}),
Sydney University Molonglo Sky Survey (SUMSS; \citealt{mauch_murphy_2003}),
and Faint Images of the Radio Sky at Twenty centimeters survey
(FIRST, \citealt{white_becker_1997}).
Whenever available we employ the given 5 GHz flux densities. However, in a few cases
we have to extrapolate the 1.4 GHz (NVSS and FIRST) or the 843 MHz (SUMSS) flux density
to the 5 GHz density by using the following relation: flux density $\propto \nu
^{\alpha_{\rm r}}$
with $\alpha_{\rm r}=-0.5$ (\citealt{kellermann_sramek_1989}).
The optical flux densities are calculated from the U.S.\ Naval Observatory survey
(USNO; \citealt{monet_etal_1998}) $B$-band magnitude at an effective wavelength
of $\sim$4100 \AA\ (\citealt{gajdoski_weinberger_1997}).

\begin{deluxetable*}{lllccccccc}
\tabletypesize{\normalsize}
\tablecaption{Observed Properties of the Most X-ray Luminous Radio-quiet QSOs in the
{\em ROSAT} Bright Survey\label{table:topten}}
\tablewidth{0pt}
\tablehead{
\colhead{Object}&\colhead{R.A.}&\colhead{Decl.}&\colhead{$B$ band}&\colhead{$R$
band}&\colhead{log($L_{\rm X}/[\rm{erg}\,\rm{s}^{-1}]$)}&\colhead{RASS
$CR$}&\colhead{$z$}&\colhead{$R$}&\colhead{Observations}\\
\colhead{Name}&\colhead{(J2000)}&\colhead{(J2000)}&\colhead{(mag)}&\colhead{(mag)}&\colhead{(0.5--2
keV)} & \colhead{(0.5--2 keV)} & \colhead{} & \colhead{$F_{\rm 5GHz}/F_{\rm Opt}$}
&\colhead{}}
\startdata
{\it RBS 1423$^{\rm a}$} & 14 44 14.7 & +06 33 07 & 16.3 & 16.7 & 46.5 &  0.11 
&0.208  & 0 & {\em XMM}        \\
{\it RBS 825$^{\rm b}$}  & 10 04 34.9 & +41 12 40 & 18.0 & 18.2 & 46.1 &  0.08 
&1.73   & 0 & {\em XMM}        \\\hline
{\bf RBS 1055}                & 11 59 41.0 &--19 59 25 & 16.5 & 16.1 & 45.3 &  0.19 
&0.450  & 5 & {\em XMM}        \\
RBS 1273                & 13 29 28.6 &--05 31 36 & 15.3 & 15.4 & 45.2 &  0.09 
&0.580  & 0 &            \\
{\bf RBS 320}                 & 02 28 15.2 &--40 57 11 & 15.4 & 14.9 & 45.1 &  0.10 
&0.494  & 0 & {\em XMM}        \\
{\bf RBS 1124}                & 12 31 36.5 & +70 44 14 & 15.3 & 15.4 & 45.0 &  0.45 
&0.209  & 0 & {\em XMM/Suzaku} \\
RBS 229                 & 01 40 17.0 &--00 50 03 & 16.5 & 16.3 & 45.0 &  0.17 
&0.334  & 1 &            \\
RBS 832                 & 10 08 31.5 & +46 29 53 & 19.3 & 18.5 & 45.0 &  0.14 
&0.388  & 0 &            \\
RBS 1062                & 12 03 43.2 & +28 35 57 & 17.2 & 17.7 & 45.0 &  0.12 
&0.374  & 0 &            \\
RBS 1502                & 15 28 40.7 & +28 25 29 & 16.1 & 16.6 & 45.0 &  0.08 
&0.450  & 0 &            \\
RBS 1853                & 22 20 49.7 &--31 56 54 & 17.6 & 17.2 & 45.0 &  0.08 
&0.503  & 0 &            \\
{\bf RBS 1883}                & 22 41 55.3 &--44 04 58 & 15.1 & 15.7 & 45.0 &  0.06 
&0.545  & 0 & {\em XMM}        \\
RBS 2040                & 23 43 13.5 &--36 37 54 & 15.4 & 15.8 & 45.0 &  0.05 
&0.620  & 2 &            \\
RBS 218                 & 01 35 20.2 &--29 23 58 & 16.7 & 15.9 & 45.0 &  0.04 
&0.699  & 0 &
\enddata
\tablenotetext{}{{\bf Notes.} The four objects in bold face are the sources discussed in this paper.}
\tablenotetext{a}{Corrected; no longer among the most X-ray luminous (AGN at
$z=0.208$ instead of an AGN at $z=2.262$, log$(L_{\rm
X}/[\rm{erg}\,\rm{s}^{-1}])=44.3$).}
\tablenotetext{b}{Corrected; no longer among the most X-ray luminous (lensed system).}
\end{deluxetable*}

\begin{deluxetable*}{lcccccccc}
\tabletypesize{\normalsize}
\tablecaption{Fit Results from Optical Spectra, Black Hole Masses, and Eddington
Ratios for the Observed RBS-QSOs\label{table:M_bh__eddington}}
\tablewidth{0pt}
\tablehead{
\colhead{Object} & \colhead{Line} & \colhead{FWHM}        & \colhead{$\lambda$} &
\colhead{$\lambda L_{\lambda}$} & \colhead{Reference}  & \colhead{$M_{\rm BH}$} &
\colhead{$L_{\rm bol}$} & \colhead{$L_{\rm bol}/L_{\rm edd}$}      \\
\colhead{Name }  & \colhead{}     & \colhead{(km s$^{-1}$)}  & \colhead{(rest-f.) (\AA)}
& \colhead{(erg s$^{-1}$)}      & \colhead{}           & \colhead{log($M_{\rm
BH}/M_{\odot}$)}& \colhead{($10^{44}$ erg s$^{-1}$)} & \colhead{      }}
\startdata
RBS 1055         &\ion{Mg}{2}     &  5540                 & 3000     &  
2.00$\times$10$^{45}$ & \cite{woo_2008}             &  8.87                  &  118 
        &       0.13 \\
RBS 320          &H$\beta$        &  2265$\pm$100         & 5100     &  
1.21$\times$10$^{46}$ & \cite{grupe_beuermann_1999} &  8.88                  &  1370
        &       1.43 \\
RBS 1124         &H$\beta$        &  4260$\pm$1250        & 5100     &  
1.26$\times$10$^{44}$ & \cite{grupe_wills_2004}     &  8.06                  &  14.2
        &       0.10 \\
RBS 1883         &H$\beta$        &  1890$\pm$200         & 5100     &  
2.58$\times$10$^{45}$ & \cite{grupe_beuermann_1999} &  8.26                  &  292 
        &       1.27
\enddata
\end{deluxetable*}

Introducing a low luminosity cut of $L_{\rm 0.5-2} = 10^{45}$\,erg s$^{-1}$,
we construct a sample of the 14 most X-ray luminous radio-quiet RBS-QSOs.
Table~\ref{table:topten} shows the main properties of the sample. The objects
are ordered in descending X-ray luminosity and count rate in the 0.5--2 keV
{\em ROSAT} band. We list the position of the optical counterpart (taken from
\citealt{schwope_hasinger_2000}), the USNO $B$- and $R$-band magnitude,
the RASS 0.5--2 keV luminosity (corrected for Galactic absorption), the RASS count rate
in the hard {\em ROSAT} band (0.5--2 keV), the redshift of the optical counterpart,
the 5 GHz to $B$-band flux density ratio $R$ (a value of $R=0$ means that the
object was not detected as a radio source), and the instrument which observed the
corresponding object. All RBS objects listed in Table~\ref{table:topten} have optical
counterparts with broad emission lines in their spectra (type I AGN). The redshifts of
our RBS sources are based on the values listed in the NED, since it contains the latest
available information on the objects.

The first two most X-ray luminous radio-quiet RBS-QSOs are RBS 1423
with $L_{\rm 0.5-2} = 10^{46.5}$ erg s$^{-1}$ and RBS 825
($L_{\rm 0.5-2} = 10^{46.1}$ erg s$^{-1}$). However, RBS 1423, whose
identification with a $z=2.26$ QSO was somewhat uncertain, indeed
was identified using the {\em XMM-Newton} position with a low luminosity
($L_{\rm 0.5-2}=2.1\times 10^{44}$ erg s$^{-1}$)
$z=0.208$ QSO with a significantly broadened ionized Fe K$\alpha$ line
(\citealt{krumpe_lamer_2007b}). Object RBS 825 turned out to be a quadruple
QSO lens system with one of the largest known separations
(\citealt{inada_oguri_2003}). Hence, its observed flux is significantly
enhanced by lensing and its intrinsic luminosity much lower
(\citealt{lamer_schwope_2006}).

Removing the exceptional objects RBS 1423 and RBS 825, the most X-ray
luminous radio-quiet RBS-QSO RBS 1055 is only the 54th most X-ray luminous
object within all RBS entries (20th most X-ray luminous when excluding BL Lac
objects). Rephrasing this interesting result, no radio-quiet QSO with
log$(L_{\rm 0.5-2}/[\rm{erg}\,\rm{s}^{-1}])>45.3$ is found in this flux limited
all-sky survey.
In contrast, the three most X-ray luminous radio-loud RBS-QSOs (RBS 717,
RBS 315, and RBS 1788) have $L_{\rm 0.5-2} = 10^{47}$ erg s$^{-1}$ and $R>2000$.
This suggests that the jet in radio-loud QSOs or jet-related changes in the
accretion disk
significantly boost the X-ray luminosity.

We have observed two of the most X-ray luminous 
radio-quiet RBS-QSOs which have high 0.5--2 keV RASS count rates 
(RBS 1055 and RBS 1124). We combine those with two archival data sets (RBS 320 and 
RBS 1883) to create a subsample of four objects that are discussed in this paper. 

%%%%%%%%%%%%%%%%%%%%%%%%%%%%%%%%%%%%%%%%%%%%%%%%%%%%%%%%%%%%%%%%%%%%%%%%%%%%%%%%%%%%%%%%%%%%%%%%%%%%%%%%%%%%%%%%%%%%%%%%%%%%%%%%%%%%%%%%%

\section{Black Hole Mass Estimates from Optical Spectra}
All four RBS-QSOs analyzed in this study have been spectroscopically observed in the
optical.
Over the last decade reverberation mapping of AGN broad emission lines established a
simple
empirical method to estimate black hole masses of a large population of AGN by
measuring the
FWHM and the luminosity which are accessible in a single spectrum (size--luminosity
relation;
see, e.g., \citealt{kaspi_smith_2000}; \citealt{vestergaard_peterson_2006}).
In this section, we give mass estimates of the central
black hole and the Eddington ratio $L_{\rm bol}/L_{\rm edd}$ for each of our objects
based on published reverberation mapping (RBS 1055, also known as CTS 306;
\citealt{woo_2008}) and
published line property and luminosity measurements from optical spectra (RBS 320,
RBS 1124, RBS 1883).

Table~\ref{table:M_bh__eddington} lists the FWHM values and the related continuum
luminosity which are used
to infer black hole masses, the reference of these values, and the derived quantities.
\cite{grupe_beuermann_1999} give only the values for FWHM of H$\beta$, but not the
$\lambda L_{\lambda}$
continuum luminosity. We used their optical spectra of RBS 320 and RBS 1883
(\citealt{beuermann_thomas_1999}; H.-C. Thomas, private communication)
to compute the relevant fluxes which were corrected for Galactic extinction. The
$M_{\rm BH}$ estimates
are based on the empirical formula from \cite{mcgill_woo_2008}. The bolometric
luminosities are estimated by
multiplying a factor of 5.9 and 11.3 (\citealt{mclure_dunlop_2004}) to the 3000 \AA\
and 5100 \AA\
continuum luminosity, respectively. The Eddington ratio $L_{\rm bol}/L_{\rm edd}$ is
the ratio between
the bolometric and the Eddington luminosity ($L_{\rm edd}= 1.26 \times 10^{38}
M_{\rm BH}/M_{\odot}$ erg s$^{-1}$).
Note that \cite{grupe_wills_2004} estimate the
bolometric luminosity from a combined fit to optical-UV and X-ray data to be $L_{\rm bol} = 3.4
\times 10^{45}$ erg s$^{-1}$. \cite{grupe_komossa_2010} list updated 
values for RBS 1124 based on simultaneous optical-to-X-ray observations with 
{\em Swift} and find an accretion rate relative to Eddington $L_{\rm bol}/L_{\rm edd}$=0.63.
However, we use the correction factor to derive an estimate of the bolometric
luminosity to treat
all sources in a consistent way.

%%%%%%%%%%%%%%%%%%%%%%%%%%%%%%%%%%%%%%%%%%%%%%%%%%%%%%%%%%%%%%%%%%%%%%%%%%%%%%%%%%%%%%%%%%%%%%%%%%%%%%%%%%%%%%%%%%%%%%%%%%%%%%%%%%%%%%%%%%

\begin{deluxetable*}{lcccccccc|cc}
\tabletypesize{\normalsize}
\tablecaption{{\em XMM-Newton} Observations of the Most Luminous Radio-quiet QSOs in
the {\em ROSAT} Bright Survey\label{table:observed}}
\tablewidth{0pt}
\tablehead{
\colhead{} & \colhead{Date}& \colhead{} & \colhead{} & \colhead{WM}      &
\colhead{Filter}     & \colhead{Net Exp.}     & \colhead{\# Photon} & \colhead{Count
Rate} & \colhead{$f_{\rm X}$} & \colhead{$L_{\rm X}$}   \\
\colhead{Object} & \colhead{(Year)}   & \colhead{ObsID} & \colhead{$N_{\rm H,GAL}$}  &
\colhead{pn} & \colhead{pn}     & \colhead{pn (ks)} &  \colhead{pn} &  \colhead{pn
(ct s$^{-1}$)} & \colhead{(0.5--2 keV)} & \colhead{(0.5--2 keV)} \\
\colhead{Name}   &\colhead{(MM-DD)} &\colhead{}    & \colhead{($10^{20}$\,cm$^{-2}$)}
& \colhead{MOS}    & \colhead{MOS}& \colhead{MOS (ks)} & \colhead{MOS} &
\colhead{MOS (ct s$^{-1}$)} & \colhead{(2--12 keV)} & \colhead{(2--12 keV)}}
\startdata
RBS 1055      & 2008  & 0555020201 & 4.12 &  PFW  & T &  21.3       &       42320 &
1.94$\pm$0.01     & 2.0  & 15    \\
              & 06-11 &            &      &  PFW  & T &  24.8       &       13904 &
0.545$\pm$0.005   & 4.1  & 28    \\
RBS 320       & 2004  & 0200480101 & 2.20 &  PFW  & T &  26.7       &       42533 &
1.57$\pm$0.01     & 1.2  & 16    \\
              & 12-07 &            &      &  PFW  & T &  31.3       &       10833 &
0.339$\pm$0.003   & 0.93 & 9.6   \\
RBS 1124      & 2008  & 0555020101 & 1.69 &  PFW  & T &  15.5       &       54901 &
5.36$\pm$0.02     & 4.6  & 5.8   \\
              & 05-15 &            &      &  PPW3 & T &  21.1       &       28701 &
1.44$\pm$0.01     & 7.1  & 8.9   \\
RBS 1883      & 2003  & 0153220101 & 1.82 &  PFW  & M &   0.5       &         422 &
0.70$\pm$0.04     & 0.57 & 11    \\
              & 05-17 &            &      &  PFW  & H &   5.2       &         670 &
0.118$\pm$0.005   & 0.40 & 5.6
\enddata
\tablenotetext{}{Abbreviations: WM -- window mode of the different detectors, PFW --
prime full window mode, PPW3 -- prime partial W3 mode (to decrease
                 the risk of photon pile-up), T -- thin filter, M -- medium filter,
H -- thick filter, Net Exp. -- net exposure time (average MOS time
                 for each MOS detector), \# Photon -- source photon number in the
0.2--12 keV band. Count rates in the energy band of 0.2--12 keV.
                 Intrinsic, Galactic absorption-corrected fluxes (in units
                 of $10^{-12}$ erg cm$^{-2}$ s$^{-1}$) and rest-frame luminosities (in units of
$10^{44}$ erg s$^{-1}$) are based on the best-fit model of
                 each object.}
\end{deluxetable*}

%%%%%%%%%%%%%%%%%%%%%%%%%%%%%%%%%%%%%%%%%%%%%%%%%%%%%%%%%%%%%%%%%%%%%%%%%%%%%%%%%%%%%%%%%%%%%%%%%%%%%%%%%%%%%%%%%%%%%%%%%%%%%%%%%%%%%%%%%%

The values for $M_{\rm BH}$, $L_{\rm bol}$, and $L_{\rm bol}/L_{\rm edd}$ in
Table~\ref{table:M_bh__eddington} are subject to large uncertainties and should be
understood as
approximate quantities only. The super-Eddington ratios of RBS 320 and RBS 1883
prompt one to investigate
the properties of the H$\beta$ lines in more detail. RBS 320 has to be modeled with
a broad and narrow Gaussian line profile. Considering only the broad line profile
with an instrumentally
corrected FWHM$ = 2900 \pm 200$ km s$^{-1}$ leads to an upper limit of the black hole
mass
log($M_{\rm BH}/M_{\odot}$) $<$ 9.09 and a lower estimate of the Eddington ratio
$L_{\rm bol}/L_{\rm edd}$ $>$0.88.
In the case of RBS 1883, the red wing of H$\beta$ falls in the atmospheric $A$-band
making the
modeling of the line even more difficult. Correcting for the $A$-band absorption, we
only needed one
broad line Gaussian profile with an instrumentally corrected FWHM$ = 3000 \pm 300$ km
s$^{-1}$ to fit the
shape of the H$\beta$ line. This results in log($M_{\rm BH}/M_{\odot}$) = 8.66 and
an Eddington ratio $L_{\rm bol}/L_{\rm edd}$=0.51.

Despite all systematics and uncertainties in the estimate, the extremely high X-ray
luminosities in these
radio-quiet QSOs originate in a combination of very high black hole masses and very
high
Eddington ratio, but they are not extreme (e.g., \citealt{schulze_wisotzki_2010};
\citealt{steinhardt_elvis_2010}).

\section{X-ray Data Reduction}
\label{reduction_analysis}

All {\em XMM-Newton} data are processed with {\tt SAS} version 7.1.0
(Science Analysis Software, \citealt{gabriel_denby_2004}) package,
including the corresponding calibration files.
The tasks {\tt emproc} and {\tt epproc} are used for generating linearized
photon event list from the raw EPIC data. The RGS data are reduced by the
task {\tt rgsproc}. The data are cleaned for times of
high particle backgrounds. We investigate the astrometrically
corrected EPIC coordinates to verify the corresponding optical
counterpart. All identifications are confirmed by our {\em XMM-Newton} data.

We make use of the Pipeline Processing System (PPS) light curves. The light
curves are background, dead time, and mirror vignetting corrected and
are extracted in an energy range of 0.2--12 keV. We rebin the light curves to
bins of 1000 s. For RBS 1883 only the MOS light curves are considered, because
of the high background count rate in the pn detector.

The X-ray spectra are extracted with the task {\tt especget}. For
the pn detector we use box-shaped background regions at
roughly the same detector $y$-position as the circular source region is
extracted. The pn background regions are located on the same chip as the
source region, except for RBS 1124 where we are forced to use the adjacent
chip due to the large covered area of RBS 1124's source region.
For the MOS detectors
we use annulus-shaped background regions around the source on the same
CCD. Additional sources in the background regions are excluded. Furthermore,
we follow the recommended flag selection of the macros {\tt XMMEA\_EP} and
{\tt XMMEA\_EM}. The low energy cut-off is set to 0.2 keV, and the high energy
cut off is fixed to 12 keV. Using the task {\tt epatplot}, we verify that
the effect of photon pile-up is negligible in all sources, except RBS 1124.
We select X-ray events corresponding to pattern 0--12 (single and doubles) for the MOS
detectors and 0--4 events for the pn detector, except for RBS 1124 where we only
select single events for the pn detector. 
The count rates of RBS 1124 (Table~\ref{table:observed}), which is the
flux-brightest radio-quiet RBS-QSO, are close to the count rate where
photon pile-up is expected (pn: $\sim$6 ct s$^{-1}$; MOS operated in large 
window-prime partial w3 mode: $\sim$1.8 ct s$^{-1}$; {\em XMM-Newton} Users
Handbook\footnote{\tt
http://xmm.esac.esa.int/external/xmm\_user\_support/documentation/uhb/index.html}).
The MOS detectors are not affected by photon pile-up. However, the pn detector shows
signs of pile-up. Therefore, we follow the recommended procedure of defining the source
region as an annulus to exclude the central region ($r=2\farcs5$) with the highest
pile-up probability.
Furthermore, we only use the pn pattern 0 spectrum which is less sensitive to pile-up
than the pattern 1--4 spectrum. The resulting 0.2--12 keV count rate for the pn detector
is therefore $cr_{\rm pn}= 3.49 \pm 0.02$ ct s$^{-1}$.

The spectra are binned to a minimum of 20 counts
per bin to apply the $\chi^2$-minimization technique. The resulting X-ray spectra
are fitted with {\tt Xspec} version 12.5.0 (\citealt{arnaud_1996}). We use the cosmic
abundances according to \cite{wilms_allen_2000} and the photoelectric absorption
cross sections provided by \cite{verner_ferland_1996}. The Galactic hydrogen column
densities in the line of sight to the objects (Table~\ref{table:observed}) are
derived from the
$N_{\rm H}$-calculator\footnote{{\tt heasarc.gsfc.nasa.gov/cgi-bin/Tools/w3nh/w3nh.pl}}
which is based on \cite{dickey_lockman_1990}.

%%%%%%%%%%%%%%%%%%%%%%%%%%%%%%%%%%%%%%%%%%%%%%%%%%%%%%%%%%%%%%%%%%%%%%%%%%%%%%%%%%%%%%%%%%%%%%%%%%%%%%%%%%%%%%%%%%%%%%%%%%%%%%%%%%%%%%%%%%
\subsection{RBS-QSOs Light Curves}

To investigate possible variability of the sources which may be related
to spectral variability, we first study the PPS light curves of
our RBS-QSO sources. The visual inspections verify that for RBS 1055,
RBS 1124, and RBS 1883 no significant variability is present.

In the case of RBS 320 the average
count rate stays nearly constant up to the middle of the observation and then
increases almost monotonically. At the end of the observation the count rate is
$\sim$20\%
higher than the average count rate at the beginning of the observation (the background
is not showing this trend). The statistical 1$\sigma$ fluctuations are on the order
of 3\%.

To investigate spectral variability, we extract two spectra for RBS 320, one
covering the first two-thirds of the
observation, the other one the last third. We apply the best-fit time-average
model to each sub-spectrum (2 power laws + Gaussian line, 2PLG; see
Section~\ref{rbs320}),
keeping as many parameters frozen at the time-average values as possible.
We first thaw only the normalization of the hard power law.
$\chi^2$/d.o.f. is 714/693 and 522/501 for the low and high flux spectra, respectively.
When we thaw the normalization of the soft power law as well, $\chi^2$/d.o.f.
falls in each case to 710/692 and 517/500. Consequently, the change of the
normalization
for the hard power law is more significant than for the soft power law when
considering the uncertainties. The Fe line parameters are consistent with their time
averaged values. Furthermore,
there is no significant evidence for variations in the photon index of either power
law. We conclude that the increase in the count rate of RBS 320 toward the end of
the observation is consistent with a change in power-law normalization mainly
in the hard component, consistent with the notion that the physical properties
of the corona (e.g., temperature) are likely not changing. The analysis of the
time-averaged spectrum over the whole observation is therefore justified.

%%%%%%%%%%%%%%%%%%%%%%%%%%%%%%%%%%%%%%%%%%%%%%%%%%%%%%%%%%%%%%%%%%%%%%%%%%%%%%%%%%%%%%%%%%%%%%%%%%%%%%%%%%%%%%%%%%%%%%%%%%%%%%%%%%%%%%%%%%
\section{Spectral Analysis}
\subsection{X-ray Spectra of RBS 1055}

\begin{figure}[t]
  \centering
  \includegraphics[width=5.1cm,clip=,angle=-90]{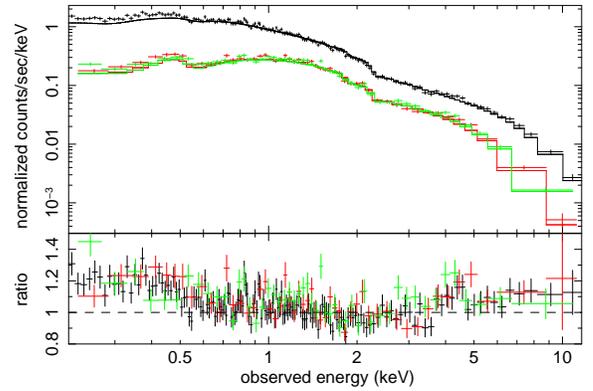}
      \caption{{\em XMM-Newton} EPIC spectra of RBS 1055 in the observed frame. The
                 spectra are
                 fitted with a single power law (1PL) to data in the 1--12 keV range.
                 At energies below 0.5 keV a soft excess is visible. A fluorescent
                 6.4 keV Fe K$\alpha$ line will be shifted to 4.4 keV at $z=0.450$.}
         \label{rbs1055_single}
\end{figure}

We start with the simplest modeling of the 0.2--12 keV data of the
most X-ray luminous radio-quiet RBS-QSO: a simple power
law (henceforth called the ``1PL'' model); this and all 
models discussed herein are corrected for Galactic absorption
(see Table~\ref{table:observed}). For all objects, studied in this paper,
no significant evidence for neutral absorption in excess of the Galactic
column density is found.
As the 1PL model is a poor fit ($\chi^2$/d.o.f. = 1413/1251),
we then try two power laws, one modeling the soft,
the other the hard energy range (henceforth called the ``2PL'' model).
$\chi^2$/d.o.f. improves to 1288/1249. The only
residuals left are positive residuals around 6.4 keV in rest frame (4.4 keV in
observed frame as shown in Figure~\ref{rbs1055_single}). We cannot successfully
model the soft excess using a blackbody component (henceforth called the ``PLBB''
model; $\chi^2$/d.o.f. = 1319/1249,
$k_{\rm B}T=0.11\pm 0.01$ keV, $\Gamma=1.73 \pm 0.01$). We return to the 2PL
and add a Gaussian fixed at 6.4 keV (rest frame) consistent with neutral
Fe K$\alpha$ emission (henceforth called the ``2PLG'' model). The fit to the data
($\chi^2$/d.o.f. = 1266/1247) and the residuals are significantly improved.

The shape of the spectrum near the Fe line cannot be explained with alternative models
incorporating absorption. For example, we try a partial covering model incorporating
absorption by gas with a column density of $N_{\rm H} \sim 6 \times 10^{23}$ cm$^{-2}$
and a covering fraction of $\sim$30\%, but this fails to properly model the residuals
near 3--6 keV ($\chi^2$/d.o.f. = 1274/1249). Although there are no obvious strong
absorption lines or edges (e.g., \ion{O}{7}, \ion{O}{8}) present in the spectrum,
we also test full-covering and partial-covering absorption by ionized material. We use
an XSTAR\footnote{{\tt http://heasarc.gsfc.nasa.gov/docs/software/xstar/xstar.html}}
table (turbulent  velocity $v=200$ km s$^{-1}$, input continuum with a
photon index $\Gamma=2$), but in either case, there is no improvement to the fit
and the data points in the 3--6 keV band are not properly modeled.
For the full-covering absorption model by ionized material the 2$\sigma$ upper limit 
estimate for the column density is $N_{\rm H,warm}=3.9 \times 10^{21}$\,cm$^{-2}$ for 
log $(\xi$/(erg cm s$^{-1}$))$ = 1.7$. In the case of the partial-covering model the 
2$\sigma$ upper limit is $N_{\rm H,warm}=1.6 \times 10^{21}$\,cm$^{-2}$ for 
log $(\xi$/(erg cm s$^{-1}$))$ = 1.7$.
In addition, no strong Fe K edge at 7.1 keV is found.
Adding a {\tt zedge} component to the 2PL model with an energy fixed at
7.1 keV, yields an upper limit of $\tau = 0.3$ and does not improve the fit
($\chi^2$/d.o.f. = 1288/1248). All these models do not account for the positive
residuals around 6.4 keV as well as the 2PLG model does. Consequently,
we conclude that the Fe K$\alpha$ emission line is real.

Leaving the line energy as a free parameter slightly improves the fit
($\chi^2$/d.o.f. = 1264/1246; henceforth called the ``2PLG2'' model). 
We find $E_{\rm Fe,rest}=6.25 \pm 0.22$ keV and
$\sigma_{\rm Fe,rest}=0.49^{+0.24}_{-0.18}$ keV. Using the response matrix, we compute
an instrumental energy resolution of FWHM$_{\rm resol}\sim 160$ eV at the line
energy.
For all objects we will only consider a line to be significantly resolved if the
line is resolved at the 95\% (2$\sigma$) confidence level.
Since a fit with an unresolved component (Gaussian line profile with a
line centroid fixed at 6.4 keV and width $\sigma=0.001$ keV) results in a worse
fit ($\chi^2$/d.o.f. = 1277/1247) and the lower limit 95\% confidence errors of
the measured $\sigma_{\rm Fe,rest}$ does not include $\sigma=0.001$ keV,
we conclude that we have detected a resolved, neutral Fe K$\alpha$ line
with EW$=0.14^{+0.14}_{-0.10}$ keV (3$\sigma$ uncertainties) in the spectrum of RBS
1055.
We exclude the presence of an additional, unresolved component: we add a second
narrow Gaussian with
energy centroid fixed at 6.4 keV (rest frame) and $\sigma$ fixed at 1 eV, but this
yields no improvement to the fit.

The freely fitted line energy of $E=6.25$ keV and
the broad shape could potentially be caused by the presence of a Compton shoulder to
the 6.4 keV Fe line. If so, we expect a Compton shoulder intensity of 20\%--40\%
of the central Fe line, with a centroid energy of $\sim$6.3 keV (\citealt{matt_2002}).
Fixing the line energy of the primary Gaussian profile to 6.4 keV and including
an additional Gaussian profile at a fixed energy of $E=6.3$ keV with an intensity
of 30\% of the primary Gaussian and a $\sigma=0.07$ keV (\citealt{bianchi_matt_2002})
does not improve the fit. Leaving the line intensity as a free parameter causes its
intensity to tend to zero and making the primary Gaussian the only significant source
for modeling the Fe line.

We then tried to model the observed Fe line profile using
a {\tt laor} component (accretion disk around a Kerr black hole; \citealt{laor_1991}),
with an emissivity index fixed to 3, inclination to 30$\degr$, and the outer radius
to 400
gravitational radii ($r_{\rm g} = GM_{\rm BH}/c^2$). We achieved a fit virtually
identical to that using
a single Gaussian emission line (see above). The inner radius of emission is
16$^{+16}_{-7}$ $r_{\rm g}$, i.e, poorly constrained due to the weakness of the line.
Similar results are found for the high S/N spectra for RBS 320 and
RBS 1124, and this model is thus not discussed further.

For RBS 1055, 320, and 1124, we determined the 90\% confidence level upper limits of
the equivalent width of
a relativistically broadened Fe line {\textit{in addition to} a narrow profile by
first modeling the Fe line with a
Gaussian component (width $\sigma$ fixed to 1 eV) and then adding a {\tt laor}
component,
leaving only the normalization of the {\tt laor} component as a free parameter
($R_{\rm in}=1.235$ $r_{\rm g}$,
$R_{\rm in} = 400$ $r_{\rm g}$, and emissivity index 3). The resulting upper limits
are discussed in
Section~\ref{origin_of_feline}.

To test if the soft excess and the broad Fe K$\alpha$ line can be modeled in a
self-consistent manner, we employed the ionized reflection model {\tt reflion}. We
use a model of the form {\tt (kdblur*reflion)+zpo}
(\citealt{ross_fabian_2005}). Given the limited S/N of our spectra,
some model parameters have to be fixed: we constrain the inclination to be 30$\degr$,
the outer radius of the reflecting disk to 400 gravitational radii,
the emissivity index $q$ to 3, and the photon index of the reflected power law equal to
that of the illuminating power law. Two almost identical $\chi^2$-fit minima are found
(the reduced $\chi^2$ values are very similar):
one fit (REFL1) is driven by both the soft excess and the Fe line and results in a
low value of the ionization parameter; the second (REFL2) is driven by the shape of the
soft excess and results in a high value of the ionization parameter.

For REFL1, the best-fit ($\chi^2$/d.o.f. = 1272/1247) yields an Fe abundance
consistent with solar (Fe/solar $=1.4^{+0.7}_{-0.6}$), an inner radius of
$R_{\rm in}=16.0^{+16.7}_{-11.4}$ $r_{\rm g}$, and the lowest
possible value of the ionization parameter, $\xi_{\rm disk}=30^{+4}_{-0*}$ erg
cm s$^{-1}$ (best-fit value pegged at hard limit of model); $\Gamma=1.77\pm 0.01$.
The fraction of the reflected component to the total 0.2--12 keV output is 9\%.
The low ionization parameter models the Fe line well, but also causes relatively
sharp line 
features in the soft excess model. Although a higher intensity of the reflected
component would
still be in agreement with the modeling of the observed Fe line, it would also
increase the
sharp line features in the soft excess which are not present in the data. Therefore,
the best-fit is an attempt to explain both features simultaneously.

In the best-fit to REFL2, the fit ($\chi^2$/d.o.f. = 1268/1247) finds Fe/solar $=
0.2^{+0.2}_{-0.1*}$,
an inner radius of $R_{\rm in} < 6.3$ $r_{\rm g}$, and $\xi_{\rm
disk}=1500^{+700}_{-600}$
erg cm s$^{-1}$; $\Gamma=1.60^{+0.07}_{-0.09}$. The reflection model comprises 34\%
of the
total 0.2--12 keV output and the parameters are driven to describe the smooth soft
excess in the data
up to 3 keV. This reflection component has only a very minor contribution to the
high energies.
Due to the high ionization parameter, the Fe K$\alpha$ line contribution is very
broad, smeared out,
and the line intensity too low. A higher Fe abundance
is needed to improve the modeling of the Fe K line profile. However, this would cause a
change in the shape of the soft excess around 0.7--0.9 keV in the rest frame,
the energy range associated with
Fe L emission, because the Fe K and Fe L line intensities are connected.
Consequently, this model is also
a compromise between accounting for the soft excess and representing the Fe line.
Ignoring the
energies around the Fe line still results in two equally good fit minima except that
both are now found with the lowest Fe abundances possible.

Leaving the emissivity index $q$ free to vary in both models leads to a somewhat
unconstrained emissivity
index and improves the fit only marginally. We keep the values of the emissivity
index frozen at 3 unless a
free index results in improving the fit significantly (see Table~\ref{reflion}).
These values may not necessarily
be representative of the physical/intrinsic emissivity indices, but such values are
required to yield a soft excess with a
smooth shape and thus achieve a good fit to the data.

The blurred reflection model can only successfully model Fe lines that originate in
close proximity to the
supermassive black hole and are relativistically broadened features. Therefore, we
consider an additional narrow Fe line component
with $E$ fixed at 6.4 keV and $\sigma$ fixed at 0.001 keV to model emission
from gas located at large distances from the central region.
This yields REFL1G and REFL2G, which differ from REFL1 and REFL2, respectively, by the
addition of the narrow Gaussian line component.
The residuals around 6.4 keV (rest frame) are better modeled in both cases (up to
$\Delta \chi^2=-6$; Table~\ref{reflion}),
and we consider REFL2G to be the best-fit model for this spectrum.

\begin{deluxetable*}{cccccc}
\tabletypesize{\normalsize}
\tablecaption{Spectral Power-Law Fits for RBS 1055, RBS 320, RBS 1124, and RBS
1883\label{powerlaw}}
\tablewidth{0pt}
\tablehead{
\colhead{Model} & \colhead{ $\Gamma$} & \colhead{$E_{\rm Fe,rest}$} &
\colhead{$\sigma_{\rm Fe,rest}$} & \colhead{EW$_{\rm Fe,rest}$} &
\colhead{$\chi^2$/d.o.f.}\\
\colhead{     } & \colhead{         } & \colhead{    (keV)        } & \colhead{      
(eV)          } & \colhead{      (eV)        } & \colhead{}}
\startdata
\multicolumn{6}{c}{{\bf RBS 1055}}\\
1PL    & 1.79$\pm$0.01                  &    --    &   --    &   --                   &       1413/1251\\
2PL    & 1.37$^{+0.20}_{-0.37}$ \& 2.12$^{+0.31}_{-0.17}$& --    &   --       &   --       &       1281/1249\\
2PLG   & 1.13$^{+0.38}_{-0.11}$ \& 1.98$^{+0.01}_{-0.10}$& !6.4! & 480$^{+220}_{-180}$ & $140^{+70}_{-50}$ & 1266/1247\\
2PLG2  & 1.07$^{+0.39}_{-0.09}$ \& 1.97$^{+0.25}_{-0.09}$& $6.25\pm 0.22$&490$^{+240}_{-180}$&$140^{+60}_{-50}$ & 1264/1246\\\hline
\multicolumn{6}{c}{{\bf RBS 320}}\\
1PL    & 2.69$\pm$0.01                                &    --   &   --  &  --      &       1615/810\\
2PL    & 1.91$^{+0.11}_{-0.12}$ \& 3.31$^{+0.13}_{-0.11}$ &    --   &    --  &  --      &        934/808\\
2PLG   & 1.95$\pm$0.10        \& 3.32$^{+0.11}_{-0.12}$ &  !6.4!  & 270$^{+420}_{-200}$&120$^{+110}_{-70}$& 924/806 \\
2PLG2  & 1.95$^{+0.11}_{-0.10}$ \& 3.32$\pm$0.12 &6.47$^{+0.21}_{-0.22}$&260$^{+370}_{-200}$&130$^{+90}_{-80}$& 924/805\\\hline
\multicolumn{6}{c}{{\bf RBS 1124}}\\
1PL    & 1.96$\pm$0.01 &    --            &    --           &    --           &       1660/1293\\
2PL    &$1.13 \pm 0.04$ \& $2.19 \pm 0.03$     &    --      &    --  &--      &       1525/1291\\
2PLG   &$1.15 \pm 0.05$ \& $ 2.19 \pm 0.04$& !6.4! &180$^{+320}_{-120}$& 70$^{+60}_{-40}$&  1516/1289\\
2PLG2  &$1.15 \pm 0.05$ \& $ 2.19 \pm 0.04$& $6.49 \pm 0.14$&140$^{+240}_{-140}$& 70$^{+60}_{-40}$&  1515/1288\\
                       &                                   &                 &                   &                     &\\
{\em 1PL}      &{\em 1.79$\pm$0.01} &    --            &    --      &    --           &      {\em 1122/926}\\
{\em 2PL}&{\em 1.58$^{+0.08}_{-0.10}$ \& 2.74$^{+0.29}_{-0.30}$}  &   --  & --  & --&      {\em   958/924}\\
{\em 2PLG}&{\em $1.58\pm 0.08$ \& 2.68$^{+0.40}_{-0.23}$}&{\em !6.40!}&{\em 120$^{+140}_{-120}$}&{\em 51$^{+34}_{-26}$}&{\em 942/922}\\
{\em 2PLG2}&{\em 1.60$^{+0.06}_{-0.10}$ \& 2.74$^{+0.33}_{-0.28}$}&{\em  6.40$^{+0.08}_{-0.12}$}&{\em 120$^{+150}_{-120}$}&{\em 52$^{+34}_{-29}$}&{\em  942/921}\\\hline
\multicolumn{6}{c}{{\bf RBS 1883}}\\
1PL   & 2.9$\pm$0.1                       &   --    &    --    &   --   &             180/86\\
2PL   & 1.7$\pm$0.3 \& 3.9$^{+0.5}_{-0.4}$  &   --    &    --    &   --   &             119/84
\enddata
\tablecomments{Models (explanation of models according to the {\tt Xspec} notation): 1PL -- single power law ({\tt phabs(zpo)}),  
2PL -- two power laws, one fitting the soft energy range, the other, the hard energy range ({\tt phabs(zpo1+zpo2)}), 2PLG -- two power 
laws including a Gaussian component with a line energy fixed to 6.4 keV in the rest frame ({\tt phabs(zpo1+zpo2+zgauss)}), 
2PLG2 -- same as 2PLG but line energy is allowed to vary. Frozen parameters are in exclamation marks. Energies and equivalent widths 
are given in the quasar rest frame. Uncertainties are computed for the 90\% confidence interval. For RBS 1124 we show the
results of the 0.35--10 keV {\em Suzaku} data fits in italic font, while all 0.2--12 keV {\em XMM-Newton}-based fits are in roman font.}
\end{deluxetable*}

\begin{deluxetable*}{cccccccc}
\tabletypesize{\normalsize}
\tablecaption{Results for REFL1G and REFL2G: Disk Reflection Plus a Narrow Fe Line
component\label{reflion}}
\tablewidth{0pt}
\tablehead{
\colhead{$\chi^2$/d.o.f.} & \colhead{ $\Gamma$} & \colhead{$q$} & \colhead{$R_{\rm in}$} & \colhead{$\xi$} &
\colhead{Fe/Solar} & \colhead{$\Delta\chi^2$}\\
\colhead{         } & \colhead{         } & \colhead{   } & \colhead{ ($r_{\rm g}$)} & \colhead{(erg cm s$^{-1}$)}
& \colhead{         } & \colhead{Fe Line}}
\startdata
\multicolumn{7}{c}{{\bf RBS 1055}}\\
1271/1246 & 1.77$^{+0.02}_{-0.01}$ & !3! &  15.9$^{+17.8}_{-10.5}$ & 30$^{+4}_{-0*}$  & 1.4$^{+0.5}_{-0.7}$   &  1 \\
1262/1246 & 1.60$^{+0.07}_{-0.09}$ & !3! &  2.6$^{+2.5}_{-1.4*}$  & 1600$^{+700}_{-600}$ & 0.2$^{+0.2}_{-0.1*}$ & 6 \\\hline
\multicolumn{7}{c}{{\bf RBS 320}}\\
933/804 & 2.65$^{+0.08}_{-0.05}$ & 8.4$^{+1.6*}_{-2.1}$ & 2.3$^{+0.3}_{-0.4}$&  110$^{+150}_{-80*}$ &0.1$^{+0.2}_{-0*}$&  7\\
971/804 & 2.39$^{+0.02}_{-0.03}$ & 7.2$^{+1.4}_{-0.9}$  & 1.9$^{+0.6}_{-0.3}$&  $1500\pm 200$     & $0.3 \pm 0.1$      &  10\\\hline
\multicolumn{7}{c}{{\bf RBS 1124}}\\
1541/1288 &1.94$^{+0.01}_{-0.02}$&!3!&  3.1$^{+2.5}_{-1.9*}$  & 30$^{+8}_{-0*}$ & 3.5$^{+2.5}_{-2.4}$ & 1\\
1514/1288 &1.55$^{+0.09}_{-0.14}$&!3!&  7.2$^{+2.8*}_{-4.4}$  & 9700$^{+300*}_{-1600}$ & 2.8$^{+3.9}_{-1.0}$ & 2\\
          &                    &   &      &                     &                      &        &\\
{\em 930/921} & {\em 1.81$^{+0.04}_{-0.03}$}&{\em !3!} & {\em 1.2$^{+1.7}_{-0*}$} &  {\em 40$^{+34}_{-10*}$}       & {\em 1.7$^{+0.9}_{-0.8}$} & {\em 4}\\
{\em 931/920} & {\em $1.17^{+0.10}_{-0.06}$}&{\em 8.0$^{+2.0*}_{-1.9}$}& {\em 1.2$^{+0.4}_{-0*}$}  & {\em 7000$^{+3000*}_{-5000}$} & {\em 3.3$^{+2.3}_{-2.5}$} & {\em 14}
\enddata
\tablecomments{For explanation see Table~\ref{powerlaw}. A lower/upper 90\%
confidence uncertainty marked with ``*'' symbolizes that the parameter
pegged at the lower/upper hard limit of the model. All fits use frozen parameters of: inclination angle = 30$\degr$, $R_{\rm out} = 400$ $r_{\rm g}$, 
$E_{\rm Fe,rest}= 6.4$ keV, and $\sigma_{\rm Fe,rest} = 1$ eV.}
\end{deluxetable*}

We also try to incorporate two reflection components to account for reflection in
different region: one very
close to the central engine producing the smooth shape of the soft excess, the other
far away to produce the
narrow Fe line. However, the moderate quality of the {\em XMM-Newton} data for all
objects does not allow us
to constrain crucial parameters such as the break radius (radius of inner and outer
reflection) and ionization
parameter of the inner disk.

The RGS data of RBS 1055, as well as for all other high S/N RBS-QSOs, are well
fitted with a single power
law ($\chi^2$/d.o.f. = 207/240). No significant absorption or emission features are
visible based on only
the RGS data of RBS 1055. The upper limits to the absolute values of equivalent widths
(90\% confidence level) of typical prominent narrow emission and absorption
lines such as \ion{O}{7}, \ion{O}{8}, \ion{Ne}{9} and \ion{Ne}{10} are as high as
$\sim 8$ eV. Both disk reflection scenarios fit equally well the RGS data.

Considering the best-fit model for RBS 1055 (REFL2G), we give the intrinsic
(corrected for Galactic absorption) 0.5--2 keV and the 2--10 keV fluxes of the
object as well as
the corresponding intrinsic luminosities for all objects in Table~\ref{table:observed}.

%%%%%%%%%%%%%%%%%%%%%%%%%%%%%%%%%%%%%%%%%%%%%%%%%%%%%%%%%%%%%%%%%%%%%%%%%%%%%%%%%%%%%%%%%%%%%%%%%%%%%%%%%%%%%%%%%%%%%%%%%%%%%%%%%%%%%%%%%%
\subsection{X-ray Spectra of RBS 320\label{rbs320}}

For the analysis of the RBS 320 data (and the remaining RBS objects), we follow the
same
approach as described in detail for RBS 1055. We will only summarize the most important
findings. As for RBS 1055, we also find that a 2PL model is required to
characterize the overall shape of the RBS 320 data appropriately ($\chi^2$/d.o.f. =
934/808) although positive residuals are still present at 6--7 keV (rest frame) where
the Fe K$\alpha$ line is expected (see Figure~\ref{rbs320_single}).

A PLBB model ($k_{\rm B}T = 0.09\pm 0.01$ keV, $\Gamma = 2.42\pm 0.02$) cannot fit the data
well ($\chi^2$/d.o.f. = 1099/808). The best-fit partial covering model incorporating
absorption by a gas with a column density of $N_{\rm H} \sim 2 \times 10^{23}$
cm$^{-2}$ and a covering fraction of $\sim$60\% fails to properly model the data
($\chi^2$/d.o.f. = 1071/808).
A moderately-good fit is obtained for full-covering absorption model with ionized
material ($\chi^2$/d.o.f. = 935/806). However, this fit finds a very high ionization
value (log $(\xi$/(erg cm s$^{-1}$))$ =6$, lower 90\% limit: log $(\xi$/(erg cm
s$^{-1}$))$ =3.3$)
and an unconstrained $N_{\rm H,warm}$ value. Ionization parameters log $(\xi$/(erg
cm s$^{-1}$))$ > 4$ do not
change the overall shape of the X-ray spectrum; they only include narrow H-like
absorption lines which are not seen in the data. Therefore, we also exclude for RBS 320 the
presence of a warm absorber.
An edge component with energy fixed at 7.1 keV does not improve the fit of the 2PL
model ($\chi^2$/d.o.f. = 934/807, 90$\%$ confidence upper limit on $\tau=0.19$).
Consequently, also in the case
of RBS 320 we conclude that an Fe K$\alpha$ emission line is detected.

Adding a Gaussian component to the 2PL model results in an improved fit
(2PLG, $\chi^2$/d.o.f. = 924/806, Table~\ref{powerlaw}) and yields
$\sigma_{\rm Fe,rest}=0.27^{+0.42}_{-0.20}$ keV for an Fe K$\alpha$ line with energy
centroid fixed at $E=6.4$ keV. The line is detected at more than 3$\sigma$ confidence.
Fixing $\sigma=0.001$ keV results in a fit with $\Delta\chi^2=+3$.
Following our general guideline for the detection of a resolved line,
the line is not resolved. Leaving the line energy as a free parameter (2PLG2) does not
change the fit. A Compton shoulder to the Fe line is not consistent with the data.

Applying REFL1 with an emissivity index fixed to 3
results in a moderately good representation of the data ($\chi^2$/d.o.f. = 1032/806)
when the lowest
possible ionization parameter is used ($\xi=30^{+3}_{-0*}$ erg cm s$^{-1}$). The fit
significantly
improves when thawing the emissivity index ($q=8^{+2*}_{-2}$, $\chi^2$/d.o.f. =
940/805). At $q=3$ the soft excess
is still modeled with some structure not seen in the data and peaks at around 0.6
keV (rest frame; \ion{O}{7}),
while much higher $q$ values again result in a much more smooth soft excess without
any structure and shift its peak to below 0.4 keV which is consistent with the data.
The Fe K$\alpha$ line is too weak and too broad in this model. The same trend can be
found when
we keep $q=3$ fixed and leave the outer disk radius as a free parameter. In this
case, the best-fit
($\chi^2$/d.o.f. = 943/805) results in an outer disk radius of only $R_{\rm
out}=4.1^{+0.4}_{-1.2}$ $r_{\rm g}$
to model the soft excess as smooth as possible. However, we decide to fix the outer
radius at
$R_{\rm out} = 400$ $r_{\rm g}$ and leave $q$ as a free parameter. The reflection
component
represents 21\% of the total 0.2--12 keV output.

As in the case for RBS 1055, we find a second best-fit model (REFL2) with a local
fit minimum at an ionization parameter of
$\xi = 1500^{+200}_{-300}$ erg cm s$^{-1}$ and $q$ as a free parameter, where the
reflection component represents 56\%
of the total 0.2--12 keV output. However, this fit minimum is worse ($\chi^2$/d.o.f.
= 981/805) than the
fit with the low ionization parameter (REFL1) because it contains a smaller
contribution at hard energies and leaves positive residuals around 6--7 keV.

Applying REFL1G and REFL2G by
adding a narrow Gaussian component with energy centroid fixed at $E=6.4$ keV and
with the line width fixed at $\sigma=0.001$ keV improves
each fit further (Table~\ref{reflion}). Leaving the line
energy as a free parameter only marginally improves the fit. The line has
EW$=0.05^{+0.04}_{-0.03}$ keV in REFL1G
(EW$=0.06^{+0.04}_{-0.03}$ keV for REFL2G).

The power-law fit of the RGS data ($\chi^2$/d.o.f. = 220/251) improves when we add
two Gaussian line
profiles in emission at rest-frame energies of $E=1.04\pm0.01$ keV and
$E=1.24^{+0.02}_{-0.04}$ keV ($\chi^2$/d.o.f. = 208/246).
Both lines are unresolved and have rest frame equivalent widths of EW$=10^{+2}_{-7}$
eV and EW$=7^{+7}_{-5}$ eV,
respectively. Because the fit improves by $\Delta \chi^2=-8$ and --5,
respectively, the lines are detected at the 2--3$\sigma$ confidence level.
In the case of the first line, the energy centroid is not far from
the expected line energy for \ion{Ne}{10}, 1.022 keV, but thanks to the resolution
of the RGS, this energy
is ruled out at more than 95$\%$ confidence. The origins of these
tentatively detected lines thus remain unclear.
Adding Gaussian components in emission or absorption at the rest-frame energies of
expected H or He-like metals
result in upper limits to the absolute value of equivalent widths up to 2 eV.
Both disk reflection scenarios fit the RGS data equally well.

\begin{figure}[t]
  \centering
  \includegraphics[width=5.1cm,clip=,angle=-90]{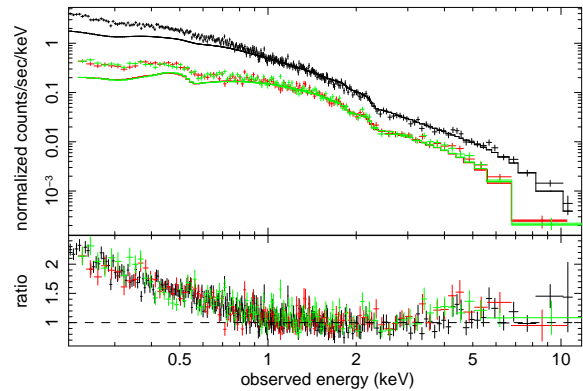}
      \caption{{\em XMM-Newton} EPIC spectra of RBS 320 in the observed frame. The
1--12 keV
                 energy range is fitted with a 1PL model. The 6.4 keV Fe
K$\alpha$
                 line will be shifted to 4.3 keV at the object's redshift of
$z=0.495$.}
         \label{rbs320_single}
\end{figure}

%%%%%%%%%%%%%%%%%%%%%%%%%%%%%%%%%%%%%%%%%%%%%%%%%%%%%%%%%%%%%%%%%%%%%%%%%%%%%%%%%%%%%%%%%%%%%%%%%%%%%%%%%%%%%%%%%%%%%%%%%%%%%%%%%%%%%%%%%%
\subsection{X-ray Spectra of RBS 1124 \label{rbs1124}}

A 2PL model again represented the data better ($\chi^2$/d.o.f. = 1525/1291)
than a 1PL model ($\chi^2$/d.o.f. = 1660/1293; see Table~\ref{powerlaw}
and Figure~\ref{rbs1124XMM_single}).
The PLBB model does not provide a better fit ($k_{\rm B}T = 0.14\pm 0.01$ keV,
$\Gamma=1.88 \pm 0.02$, $\chi^2$/d.o.f. = 1580/1291). A partial covering model with
a column density of $N_{\rm H} \sim 4 \times 10^{23}$ cm$^{-2}$ and a covering fraction of $\sim$21\%
describes the data very well ($\chi^2$/d.o.f. = 1515/1291). A full-covering and partial-covering
absorption model with ionized material can be excluded because they both require very high
column densities and ionization parameters ($N_{\rm H} \sim 1 \times 10^{24}$ cm$^{-2}$,
log $(\xi$/(erg cm s$^{-1}$))$ = 5$; see explanation for RBS 320). Furthermore, we
exclude an Fe edge component at 7.1 keV (rest frame) to account for the positive residuals
around 6 keV (rest frame) in 2PL model (90$\%$ confidence upper limit on $\tau=0.02$).

Adding a Gaussian component to a 2PL model improved the fit further by
$\Delta \chi^2 = -9$ (2PLG, $\chi^2$/d.o.f. = 1516/1289). Consequently, the line is detected at
the 3$\sigma$ confidence level. Fixing the width of the line to $\sigma=0.001$ keV
yields $\chi^2$/d.o.f. = 1521/1290. The line is resolved since the lower 2$\sigma$ limit of the line width
is not
consistent with the expected value for an unresolved line.
Leaving the line energy as a free parameter (2PLG2) only improves the fit marginally
($\Delta \chi^2 = -1$).

The self-consistent disk reflection model (REFL2) results in a very good fit of the data
(emissivity index fixed to 3, $\chi^2$/d.o.f. = 1516/1289). We find an extremely
high ionization parameter of $\xi = 9000^{+1000*}_{-2000}$ erg cm s$^{-1}$ and almost solar abundances.
The reflection component accounts for 75\% of the total 0.2--12 keV output.
With the best-fit parameters of this model, the shape of the reflection component is
very similar to a simple power-law fit except that it models a very broad and smeared out Fe line.

As for RBS 1055 and RBS 320, a second stable local $\chi^2$-minimum is found for a
fit with low $\xi$ values (REFL1, $\chi^2$/d.o.f. = 1542/1289). REFL1
introduces discrete line features to model the soft excess and the Fe line. Since
RBS 1124's soft excess is again very smooth, the reflection component can only account for 6\% of
the total 0.2--12 keV output in order to avoid introducing significant structure to the soft excess.
Furthermore, REFL1 predicts a slope which is too steep at high energies compared to the data.

Thawing the emissivity index $q$ does not affect the quality of REFL1 and REFL2
significantly and $q$ is found to be unconstrained in both fit models. Therefore, we decide to leave it
fixed to $q=3$. Considering an additional narrow Gaussian line profile (REFL1G and REFL2G) with
a fixed rest-frame energy of 6.4 keV improves both disk reflection fit models marginally. The line has an
EW$=0.03^{+0.03}_{-0.02}$ keV in REFL1G and REFL2G.

After fitting the RGS data with a single power-law model ($\chi^2$/d.o.f. =
349/349), no narrow striking spectral features are visible (upper limits to absolute values of
EWs on prominent narrow emission and absorption lines are up to
EW $\sim$ 5 eV). The RGS data are again equally well fitted by both reflection
scenarios.

\begin{figure}[t]
  \centering
  \includegraphics[width=5.1cm,clip=,angle=-90]{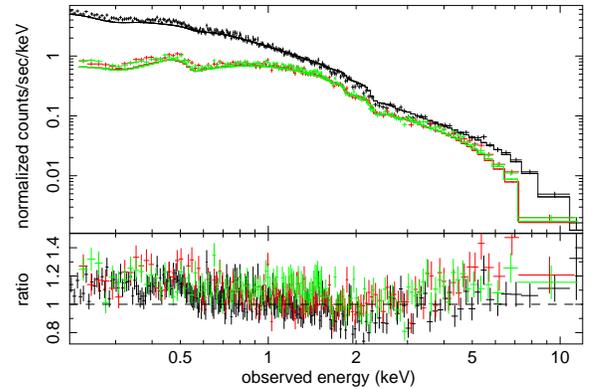}
      \caption{{\em XMM-Newton} EPIC spectra of RBS 1124 in observer frame. A 1PL 
                 is fitted to the 1--12 keV energy band. The 6.4 keV Fe K$\alpha$
                 line will be
                 shifted to 5.3 keV at the object's redshift of $z=0.209$.}
         \label{rbs1124XMM_single}
\end{figure}

%%%%%%%%%%%%%%%%%%%%%%%%%%%%%%%%%%%%%%%%%%%%%%%%%%%%%%%%%%%%%%%%%%%%%%%%%%%%%%%%%%%%%%%%%%%%%%%%%%%%%%%

\subsection{Joint Fit Analysis of {\em XMM-Newton} and {\em Suzaku} Spectra}
RBS 1124 was also observed with {\em Suzaku} on April 14--16, 2007 (ObsID 702114010)
for a duration of 138 ks (before screening for Earth occultation, etc.). The
analysis and
results of this observation are published in \cite{miniutti_piconcelli_2010}.
We include in this paper joint spectral fitting of the {\em XMM-Newton} and {\em
Suzaku} spectra
to achieve better constraints on spectral fit parameters for the Fe line compared to
{\em XMM-Newton}
alone. In addition, the two observations of RBS 1124 gives us the opportunity to
study spectral variability in one of the most
luminous radio-quiet QSOs on a 13 month time scale and with comparable S/N spectra.

\subsubsection{{\em Suzaku} RBS 1124 Data Reduction}
The data were obtained with the X-ray imaging CCDs (XIS; \citealt{koyama_Tsunemi_2007})
on board {\em Suzaku}, which consist of two front-illuminated (FI) CCD and one
back-illuminated (BI)
CCD (XIS2 was not used), as well as the
non-imaging hard X-ray detector (HXD; \citealt{takahashi_abe_2007}). RBS 1124 was
placed at the
``HXD nominal'' position. The XIS and HXD data are processed with {\em Suzaku}
pipeline processing
version 2.0.6.13. The analysis uses the HEADAS version 6.6.1.\ software.
The XIS data reduction follow the guidelines provided in the {\em Suzaku} Data
Reduction
Guide\footnote{http://heasarc.gsfc.nasa.gov/docs/suzaku/analysis/abc/abc.html}. As per
a notice for a calibration database
update\footnote{http://heasarc.gsfc.nasa.gov/docs/suzaku/analysis/sci\_gain\_update.html}
the task {\tt xispi} is run on the unfiltered events again. The events are filtered
using
standard extraction criteria (e.g., avoiding the South Atlantic Anomaly, low Earth
elevation angles, etc.).
The XIS good integration time is 86 ks per CCD.
We extract for each of the three XIS CCDs the source events by choosing a region
3$\arcmin$ in radius,
the background events by four regions 1$\farcm$5 in radius, and the $^{55}$Fe
calibration source events.
RBS 1124 is detected with a net count rate of 0.29 ct s$^{-1}$ in the XIS FI and 0.39
ct s$^{-1}$ in the XIS BI. The response (RMF) and auxiliary (ARF) file are produced
by using
the tasks {\tt xisrmfgen} and {\tt xissimarfgen}. The data for both XIS FI detectors
are co-added to form a single
spectrum using {\tt mathpha}, {\tt addrmf}, and {\tt addarf}. Finally the data are
grouped to at least 20 counts per bin.

The \ion{Mn}{1} K$\alpha_1$ and K$\alpha_2$ lines in the calibration spectrum are
fitted
by two Gaussian components, where the K$\alpha_2$ Gaussian component has half the intensity of
K$\alpha_1$.
The nominal rest-frame energies of Mn I K$\alpha_1$ and K$\alpha_2$ are 5.899 keV and
5.888 keV, respectively, with an expected intensity ratio of 2:1. The observed
K$\alpha_1$ line peaks at
$E=5.897 \pm 0.001$ keV for the added XIS FI and  $E=5.897 \pm 0.002$ keV for the
XIS BI with an upper limit up to $\sigma=16$ eV (co-added FI and BI).
The FWHM energy resolution is 149 eV for the present observation.

The 12--76 keV HXD/PIN data are also reduced following the {\em Suzaku} guidelines.
As in \cite{miniutti_piconcelli_2010},
we find that an extrapolation of the XIS spectral fit to higher energies does not
agree with the PIN data.
\cite{walton_reis_2010} explain similar {\em Suzaku} spectra (e.g., 1H~0419-577)
with two reflection disk
models: one (with {\tt kdblur}) modeling the effects of strong gravity in the
innermost region of the disk,
the other (without {\tt kdblur}) accounting for reflection at far distances.
Although modeling the data better
than a 2PL fit, the proposed models, including splitting the disk into an inner and
outer region,
still leave strong positive residuals at energies below 20 keV (observed frame).
Since the origin of the observed
steep excess in the PIN data below 20 keV (observed frame) still remains unclear
(likely an additional variable hard
X-ray source in the HXD/PIN field of view or instrumental issues), we decide not to
use the PIN data.

\subsubsection{{\em Suzaku} X-ray Spectra of RBS 1124}
\cite{miniutti_piconcelli_2010} focus on fitting the {\em Suzaku} RBS 1124 data
with the self-consistent disk reflection model {\tt reflion} and a partially
covering model. We briefly summarize our fit results which also includes detailed power-law fits on
the {\em Suzaku} 0.3--10 keV spectrum of RBS 1124. We fit the spectrum over 0.3--9 keV for the XIS BI
and 0.4--11.5 keV for the co-added XIS FI. Furthermore, we ignored energies in the range 1.75--1.90
keV to avoid \mbox{calibration} 
uncertainties associated with the instrumental Si K-edge. Leaving the normalization
difference between the XIS FI and BI as a free parameter to account for instrumental
cross-calibration improves all fits (e.g., 1PL by $\Delta \chi^2=-15$). We find a
normalization constant between both instrument types of approximately 0.98.

A 2PL model fits also the {\em Suzaku} data better than a 1PL model
(Table~\ref{powerlaw}, $\chi^2$/d.o.f. = 958/924 versus $\chi^2$/d.o.f. = 1122/926).
The PLBB model is a very good representation of the data ($\chi^2$/d.o.f. = 947/924,
$\Gamma=1.72\pm 0.02$, $k_{\rm B}T = 0.14\pm 0.01$ keV). A partial-covering absorption model
with neutral material does not explain the data well ($\chi^2$/d.o.f. = 968/924,
$\Gamma=1.94\pm 0.03$,
$N_{\rm H} \sim 2 \times 10^{23}$ cm$^{-2}$, and a covering fraction of $\sim$30\%).
There are no obvious narrow discrete absorption features seen that would
suggest the presence of an ionized absorber as well. We return
to the 2PL fit and add a Gaussian line profile. The 2PLG model improves the fit ($\chi^2$/d.o.f.\ =
942/921) and an Fe line is detected at greater
than the 4$\sigma$ confidence level. Fixing the line width to
$\sigma=0.001$ keV results in $\chi^2$/d.o.f.= 944/922. The line is not resolved.

REFL1, the self-consistent disk reflection model without an additional narrow Fe
line, is a very good representation of the data ($\chi^2$/d.o.f. = 934/921, 
emissivity index frozen) when an ionization parameter of 
$\xi_{\rm disk} = 38^{+25}_{-8*}$ erg cm s$^{-1}$ is used. The reflection
component accounts for 13\% of the total 0.2--12 keV output.
We also find in the {\em Suzaku} data a second fit minimum incorporating a high
ionization parameter (REFL2, $\chi^2$/d.o.f. = 967/922, 
$\xi_{\rm disk} = 2200^{+1200}_{-800}$ erg cm s$^{-1}$ and keeping the emissivity index fixed to 3). REFL2 significantly improves 
($\Delta \chi^2$ = -22) when the emissivity
index is allowed to vary ($q=10^{+0*}_{-3}$, $\xi_{\rm disk} =
6700^{+3300*}_{-2800}$ erg cm s$^{-1}$).
This reflection model makes up 54\% of the total 0.2--12 keV output.

\begin{figure}[t]
  \centering
  \includegraphics[width=5.1cm,clip=,angle=-90]{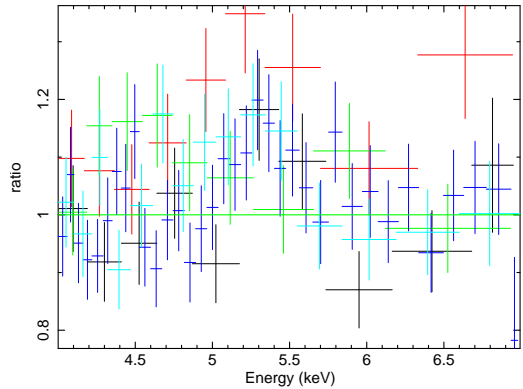}
      \caption{To illustrate the Fe K$\alpha$ fluorescence line in RBS 1124 
               we show the data/model residuals for a 2PL model fit, i.e., 
               without the Fe line modeled, to the {\em XMM-Newton} EPIC 
               (black -- pn, red -- MOS1, and green -- MOS2) and 
               {\em Suzaku} (dark blue -- XIS FI, light blue -- XIS BI) spectra.}  
         \label{join_fe_line}
\end{figure}

We then added a narrow Fe line to the reflection models (REFL1G and 
REFL2G) and find improvements to the fit compared to REFL1 and REFL2, 
respectively (Table~\ref{reflion}). Model REFL2G improves by 
$\Delta \chi^2$ = -24 when the emissivity index is left as a free 
parameter. Our fit results for model REFL1G are consistent with the 
values given in \cite{miniutti_piconcelli_2010} considering the 
uncertainties. Based on model REFL1G we derive intrinsic 
$f_{\rm 0.5-2}= 3.0 \times 10^{-12}$ erg cm$^{-2}$ s$^{-1}$
($f_{\rm 2-10}= 4.8 \times 10^{-12}$ erg cm$^{-2}$ s$^{-1}$) and
a corresponding $L_{\rm 0.5-2}= 3.8 \times 10^{44}$ erg s$^{-1}$
($L_{\rm 2-10}= 5.8 \times 10^{44}$ erg s$^{-1}$).  

The total 0.5--2 keV flux and total 2--10 keV flux both increased 
between the {\em Suzaku} and {\em XMM-Newton} observation, by roughly 
60\% and 35\%, respectively. In the context of the 2PL model, there is 
evidence for both the 1 keV normalization and photon index of both power-law 
components (soft and hard) to vary between the observations.
The soft excess contribution according to a phenomenological blackbody model
(computing the contribution of the blackbody model to the total 0.5--2 keV
flux) has decreased from ($9\pm 1$)\% in the {\em Suzaku} observation
to ($5\pm 1$)\% in the {\em XMM-Newton} observation 13 months later. Because 
the soft excess is observed to vary, an origin in a very extended or distant 
region (greater than 13 light months from the central X-ray source) is ruled 
out. The detailed relation between the origins and interactions of the soft 
excess and the hard power law for RBS 1124 remain unclear. However, the parameters 
for the Fe line profile do not vary between the two observations (line parameters 
are consistent within the 1$\sigma$ uncertainties), and so we can do a joint 
fit to achieve better constraints on the Fe line.

\subsubsection{Fe Line Constraints Through a Joint Fit}

For the joint fit we use the same energy ranges that we used for the
single spectral fits. We apply the 2PLG model to the data. For the
{\em Suzaku} data we keep the BI/FI normalization constant frozen
at $c=0.98$. Tying the slopes of the soft and hard power laws 
for both observation and only using different normalizations yield 
a very bad fit with very significant residuals. Consequently, 
we leave both the 1 keV normalization and the photon indices 
of the soft and hard power law free to vary between the {\em Suzaku} and 
{\em XMM-Newton} observations. Allowing the line width and the line
energy to vary between both observations does not improve the fit further.

We derive an Fe line width of $\sigma_{\rm Fe,rest}=0.14^{+0.10}_{-0.09}$ keV 
from the joint fit (Figure~\ref{join_fe_line}). If we omit the line component,
the fit worsens by  $\Delta$$\chi^2=24$, while fixing the line width to 1 eV 
worsens the fit by $\Delta$$\chi^2=8$. Consequently, using a joint fit to 
both observations, we detect the Fe line with a much higher significance 
than in the individual observations. The line is resolved and has an intensity 
of $I= (5.2 \pm 2.1) \times 10^{-6}$ ph cm$^{-2}$ s$^{-1}$.

%%%%%%%%%%%%%%%%%%%%%%%%%%%%%%%%%%%%%%%%%%%%%%%%%%%%%%%%%%%%%%%%%%%%%%%%%%%%%%%%%%%%%%%%%%%%%%%%%%%%%%%%%%%%%%%%%%%%%%%%%%%%%%%%%%%%%%%%%%
\subsection{X-ray Spectra of RBS 1883}

The observation of RBS 1883 suffers heavily from very high background count rates 
throughout almost the entire observation. Although the low number of counts does 
not allow us to study RBS 1883 as detailed as the other RBS QSOs presented in this 
paper, a 2PL model fits the data much better than a 1PL ($\chi^2$/d.o.f. = 180/86
versus 119/84; see Table~\ref{powerlaw} and Figure~\ref{rbs1883_single}). A PLBB model
results in a fair fit of the data ($k_{\rm B}T = 0.13 \pm 0.01$; $\chi^2$/d.o.f. = 124/84).

A partial covering model improves the fit ($\chi^2$/d.o.f. = 113/84) by finding a
very high covering fraction of (82$\pm$6)\% and 
$N_{\rm H,cov} = (1.4^{+0.6}_{-0.4}) \times 10^{23}$ cm$^{-2}$. The S/N of the data
does not allow us to test for a warm absorber in RBS 1883; the ionization parameter
and $N_{\rm H,warm}$ stay unconstrained. For the same reason we are not able to
apply a reflecting disk model to the data.

Although no emission line features, e.g., an Fe K$\alpha$ line, are visible in the
spectra, we included a Gaussian component fixed to an energy of 6.4 keV (rest frame) 
to give upper limits on the Fe K$\alpha$ emission line. For an unresolved Fe line 
(width $\sigma = 0.001$ keV), we find an 90\% confidence level upper limit on the 
EW$ < 0.66$ keV (EW$ <0.84$ keV for a resolved Fe line with $\sigma$ fixed to 0.4 keV).

\begin{figure}[t]
  \centering
  \includegraphics[width=5.1cm,clip=,angle=-90]{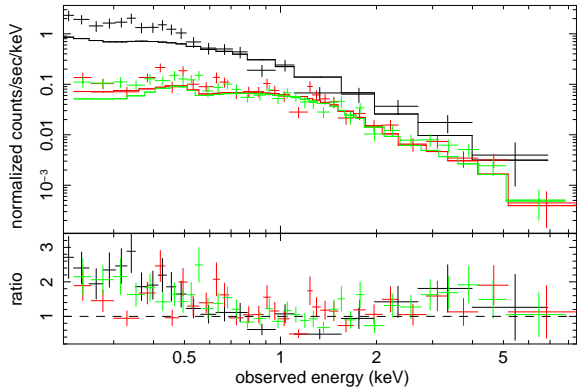}
      \caption{{\em XMM-Newton} EPIC spectra of $z=0.545$ RBS 1883 in observer
               frame. A 1PL is fitted to the 0.5--12 keV energy band.}
         \label{rbs1883_single}
\end{figure}

%%%%%%%%%%%%%%%%%%%%%%%%%%%%%%%%%%%%%%%%%%%%%%%%%%%%%%%%%%%%%%%%%%%%%%%%%%%%%%%%%%%%%%%%%%%%%%%%%%%%%%%%%%%%%%%%%%%%%%%%%%%%%%%%%%%%%%%%%%
\section{Optical-to-X-ray spectral Indices}

The optical monitor (OM) onboard {\em XMM-Newton} provides
UV photometric data for all objects taken simultaneously with the X-ray
data and therefore enables us to calculate optical-to-X-ray spectral energy
distributions ($\alpha_{\rm ox}$) without the uncertainties introduced by 
source variability. 
The OM source lists produced by the XMM-SAS pipeline contain
calibrated source magnitudes in the AB system that are also corrected
for the long term degradation of the OM.  
In addition, we de-redden the UV data using the maps by 
\cite{schlegel_Finkbeiner_1998} and the wavelength dependencies by 
\cite{cardelli_clayton_1989}. 

The four OM observations of the RBS-QSOs were taken with different 
filter combinations. To derive the $k$-corrected rest-frame luminosity densities at 2500 \AA ,
we use the mean UV continuum slope $\alpha_{\rm uv}=-0.44$ (frequency index) of the SDSS QSO spectra 
(\citealt{vandenberk_richards_2001}). If we have more than one magnitude measurement 
for an individual object, we select the filter observation which is less 
affected by a strong UV emission line. 
The only OM detection of RBS 320 in filter UVM2 coincides with the \ion{C}{4} 
line in the object and the derived values should be considered with care. 
The rest-frame luminosity densities at 2 keV is computed from the 
0.5--2 keV rest-frame luminosities given in Table~\ref{table:observed}
and the photon indices measured in the same energy range (Table~\ref{table:compare}).
Considering the redshift of our objects, the 2 keV rest-frame 
luminosity density will be redshifted into the 0.5--2 keV band. 
The measured magnitudes and the derived quantities are listed in 
Table~\ref{table:alphaox}.

\begin{deluxetable}{lcccc}
\tabletypesize{\normalsize}
\tablecaption{Magnitudes, Rest-frame Luminosity Densities, $\alpha_{\rm ox}$\label{table:alphaox}}
\tablewidth{0pt}
\tablehead{
\colhead{Object} & \colhead{AB Magnitude/}   & \colhead{log$l_{2500{\rm\,\AA}}$}  & \colhead{log$l_{\rm 2\,keV}$} & \colhead{$\alpha_{\rm ox}$} \\
\colhead{Name }  & \colhead{Filter}       & \colhead{}                      &\colhead{}               & \colhead{}}
\startdata
RBS 1055         & 18.18$\pm$0.02/UVW1$^*$&  30.11   & 27.41   & 1.04 \\
                 & 17.26$\pm$0.03/UVM2    &          &         &      \\    
RBS 320          & 15.44$\pm$0.01/UVM2    &  31.25   & 27.15   & 1.26 \\
RBS 1124         & 16.44$\pm$0.01/UVW1    &  30.07   & 26.94   & 1.20 \\
RBS 1883         & 16.33$\pm$0.01/UVW1    &          &         &      \\
                 & 16.29$\pm$0.02/UVW2$^*$&  30.98   & 26.91   & 1.57         
\enddata
\tablenotetext{}{The magnitudes with the superscript ``*'' are less affected by strong UV emission lines and are used for the 
                 computation of the rest-frame luminosity at 2500 \AA\ and $\alpha_{\rm ox}$.
                 The effective wavelengths for the filters are: UVW2 -- 2120 \AA , UVM2 -- 2310 \AA , and UVW1 -- 2910 \AA .
                 The rest-frame luminosity densities $l_{2500{\rm\,\AA}}$ and $l_{\rm 2\,keV}$ are given in units of erg s$^{-1}$ Hz$^{-1}$.}
\end{deluxetable}

The derived $\alpha_{\rm ox}$ values cover the same range that is found by \cite{grupe_komossa_2010} in a sample of soft 
X-ray selected AGN drawn from the RASS survey and observed with {\em Swift}. They also observed RBS 1124 and find an 
$\alpha_{\rm ox}= 1.23$. Considering this {\em Swift} sample, RBS 1055 yields an $\alpha_{\rm ox}$ value on the low edge of the 
distribution. 
Our values are also consistent with the sample by \cite{anderson_margon_2007, anderson_voges_2003} 
who spectroscopically identified 7000 RASS AGN from the Sloan Digital Sky Survey (data release 5).

%%%%%%%%%%%%%%%%%%%%%%%%%%%%%%%%%%%%%%%%%%%%%%%%%%%%%%%%%%%%%%%%%%%%%%%%%%%%%%%%%%%%%%%%%%%%%%%%%%%%%%%%%%%%%%%%%%%%%%%%%%%%%%%%%%%%%%%%%%

\section{Long-term Flux Variability of the RBS-QSO{\scriptsize s}}
%%%%%%%%%%%%%%%%%%%%%%%%%%%%%%%%%%%%%%%%%%%%%%%%%%%%%%%%%%%%%%%%%%%%%%%%%%%%%%%%%%%%%%%%%%%%%%%%%%%%%%%%%%%%%%%%%%%%%%%%%%%%%%%%%%%%%%%%%%

Since all the studied objects in this work have been observed with {\em ROSAT} and
{\em XMM-Newton} separated by almost two decades, both measurements allow a crude 
long-term flux variability study. The RBS contains the flux-brightest (with some 
spatial restrictions) objects from the RASS which was observed in 1990/1991 during 
the first half year of the {\em ROSAT} mission. RBS 1124 has also been observed by 
the {\em XMM-Newton} Slew Survey, {\em Suzaku}, and {\em Swift} 
(see \citealt{miniutti_piconcelli_2010}). RBS 320 was observed four years after the
{\em XMM-Newton} observation with {\em Swift} on 2008 November 7 
(target name: QSO B0226-4110; target ID: 37516). We used the online {\em Swift} XRT 
data reduction pipeline\footnote{\tt http://www.swift.ac.uk/user\_objects/}
(\citealt{evans_beardmore_2009}) to extract a $\sim$13 ks spectrum. Binning the data
to 20 counts per bin, the low S/N ($\sim$1000 counts) {\em Swift} spectrum requires 
only a single power-law fit ($\Gamma=2.58\pm 0.09$, $\chi^2$/d.o.f.\ = 41/42) with 
no intrinsic absorption. This is consistent with the 1PL fit of the {\em XMM-Newton} 
data considering the uncertainties.

\begin{figure}[t]
  \centering
 \resizebox{\hsize}{!}{
  \includegraphics[bbllx=101,bblly=372,bburx=553,bbury=700]{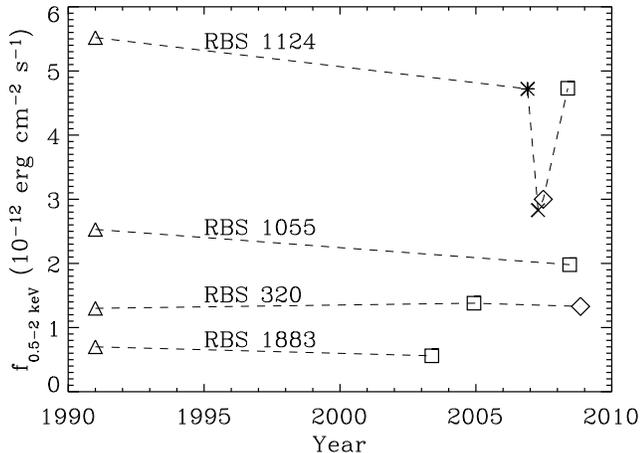}}
      \caption{Intrinsic (Galactic-absorption corrected) flux vs. time for the
               studied RBS-QSOs. The symbols represent measurements from different 
               missions: triangle -- {\em ROSAT}, square -- {\em XMM-Newton}, 
               diamond -- {\em Swift}, asterisk -- {\em XMM-Newton} Slew Survey, 
               and cross -- {\em Suzaku}.  }
         \label{RBS_lightcurve}
\end{figure}

The 0.5--2 keV {\em ROSAT} fluxes are taken from the RBS catalog 
(\citealt{schwope_hasinger_2000}). They assumed a photon index of $\Gamma=2$ to
compute count-to-energy conversion factor for the {\em ROSAT} PSPC detectors 
considering the Galactic column density toward the X-ray source. To keep possible 
flux changes due to the different flux computation methods between {\em ROSAT} (RBS)
and the measurements from other missions as low as possible (flux differences up 
to 10\%), we applied the RBS technique to all observations. 
Figure~\ref{RBS_lightcurve} shows all measured fluxes for the studied RBS-QSOs 
versus time. Except for the decrease and rise of RBS's 1124 flux between 2006 November 
and 2008 May, all objects show only flux variability on a level of 
$\sim$5\%--20\% based on the few data points available.

%%%%%%%%%%%%%%%%%%%%%%%%%%%%%%%%%%%%%%%%%%%%%%%%%%%%%%%%%%%%%%%%%%%%%%%%%%%%%%%%%%%%%%%%%%%%%%%%%%%%%%%%%%%%%%%%%%%%%%%%%%%%%%%%%%%%%%%%%%
\section{Construction of an Average X-ray Spectrum}
\label{stacked_spectrum}
We decided to construct an average X-ray spectrum for the four sources studied here.
First, the average spectrum may unveil a possible relativistic component of the Fe
K$\alpha$ emission line in the most luminous radio-quiet RBS-QSOs due to the increase 
in S/N. Secondly, the average broadband spectral properties in these objects can 
be compared to the average broadband spectral properties obtained for high-redshift,
low S/N spectra with very similar X-ray luminosities found routinely by the 
{\em XMM-Newton} and {\em Chandra} surveys (e.g., \citealt{mateos_carrera_2010};
\citealt{brusa_gilli_2005}). This will test if our most luminous radio-quiet RBS-QSO 
X-ray spectra, which allow us to determine their spectral properties directly, are 
the standard cases of luminous radio-quiet QSOs at high redshifts.

The averaging process follows \cite{corral_page_2008} with some differences.
The original process was designed to average a large number of sources to
improve the S/N, whereas here we are only averaging four sources. Therefore, the
influences of individual sources on the average spectrum is considerably higher. We 
tested the source-to-source influence on the average spectrum by generating
sets of average spectra, removing one source in each case. The results are robust,
i.e., the derived properties vary within their 2$\sigma$ limits. In particular, 
removing RBS 1055, which has the broadest Fe line, does not change the properties of 
the averaged Fe line.

We are forced to exclude the pn data for RBS 1883 since, given the small number of
counts at high energies in this case, it distorts the average spectrum shape at those
energies. The main steps of the averaging process are as follows:
We fit a 2PL model to each spectrum, instead of the simple absorbed power law as
described in \cite{corral_page_2008}, since, as we show above, this is a very good
representation of the X-ray data for the whole sample. Then we obtain the incident 
spectrum for each source, i.e., the source spectrum before entering the detector, 
by using the previously obtained model and the {\tt Xspec} command {\tt eufspec}. 
Doing so, we derive the source spectra in flux units of keV cm$^{-2}$ s$^{-1}$ keV$^{-1}$. 
Next, we de-absorb each spectrum for the Galactic absorption at each source
position and shift them to their rest frame. To combine the individual spectra, we
rescale them by forcing the rest frame 2--5 keV flux to be the same for all objects. 
Finally, we rebin the spectra to a common energy grid in a range of 1--13 keV, 
ensuring that all bins contain at least 600 real counts, and average them by using a 
standard mean. Individual errors on the real spectra are propagated as Gaussian 
along the whole averaging process. 
The significances of the individual spectral features in the final average 
spectrum are quantified by using simulations. We simulate each source 100 times using 
its best-fit model and keeping the same spectral quality and flux as for the real source. 
We apply exactly the same averaging process to the simulated spectra as we do for the 
real spectra.
From the 100 simulated spectra, we compute the confidence limits (1$\sigma$ and 2$\sigma$ 
confidence encompass 68\% and 95\% of the simulated values, respectively) and a simulated 
continuum, which account for the mixture of 2PL shapes. As already suggested by the 
individual X-ray spectra, the resulting average spectrum contains in comparison to a 2PL 
no spectral features except an Fe line which is detected above the 2$\sigma$ confidence 
limit. The Fe K bandpass is shown in Figure~\ref{stacked_Fe_line}.

\begin{figure}[t]
  \centering
 \resizebox{\hsize}{!}{
  \includegraphics[bbllx=74,bblly=368,bburx=538,bbury=700]{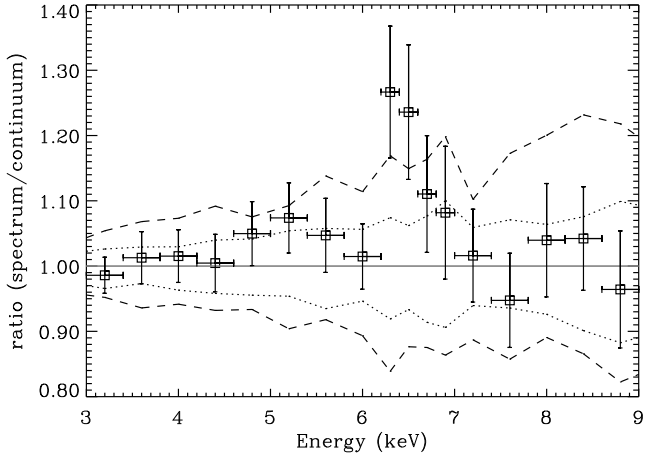}}
      \caption{Added {\em XMM-Newton} spectrum of RBS 1055, RBS 320, RBS 1124, and
               RBS 1883. The stack contains $\sim$35,000 counts in the shown 3--9 keV 
               range. The squares represent the ratio between the flux of the stacked 
               spectrum and the simulated continuum spectrum. The uncertainties of the 
               bins of the stacked spectrum are shown as error bars. The dotted lines show the
               1$\sigma$ confidence limits, while the dashed line represents the 2$\sigma$
               confidence limits. The solid gray line at a ratio of one illustrates the 
               continuum flux.}
         \label{stacked_Fe_line}
\end{figure}

Following the analysis on individual objects, we first fit a 2PL to the 1--12 keV
rest frame data. We achieve a good fit of $\chi^{2}$/d.o.f. = 22/19 
($\Gamma_{\rm soft} = 3.4^{+0.7}_{-0.5}$, $\Gamma_{\rm hard} = 1.7^{+0.1}_{-0.2}$). 
The remaining positive residuals can be well fitted by adding a Gaussian profile. 
The 2PLG model improves the fit to $\chi^{2}$/d.o.f. = 8/16 and yields 
$\Gamma_{\rm soft} = 3.4^{+0.8}_{-0.6}$, $A_{\rm soft}=0.8^{+0.3}_{-0.2}$  
ph cm$^{-2}$ s$^{-1}$ keV$^{-1}$ (normalization at 1 keV), $\Gamma_{\rm hard} = 1.7^{+0.1}_{-0.2}$, 
$A_{\rm hard}=0.7^{+0.2}_{-0.3}$ ph cm$^{-2}$ s$^{-1}$ keV$^{-1}$, 
$E_{\rm Fe,rest}=6.4\pm 0.1$ keV , $\sigma_{\rm Fe,rest}=0.2\pm 0.2$ keV,
and $I_{\rm Fe}=0.004\pm0.002$ ph cm$^{-2}$ s$^{-1}$. 
We find an averaged rest-frame equivalent width of EW$ = 0.12\pm 0.06$ keV in the 
stacked spectrum of the RBS-QSOs. The fit still leaves some tempting looking positive 
residuals around 4.5--6 keV (see Figure~\ref{stacked_Fe_line}), but adding an additional 
Gaussian profile does not yield a statistically significant improvement in the fit. 
In order to give at least an upper limit for the equivalent width of a broadened disk 
line, we add a {\tt laor} profile to the 2PLG model. We only allow the normalization of the 
{\tt laor} disk line to vary ($R_{\rm in}$ fixed to 1.235 $r_{\rm g}$) and find EW$ < 340$ eV. 

\begin{deluxetable}{ccc}
\tabletypesize{\normalsize}
\tablecaption{1PL Photon Indices for Different Redshifts of
the Averaged RBS-QSO Spectrum\label{table:shift_redshift}}
\tablewidth{0pt}
\tablehead{
\colhead{$z$} & \colhead{Photon index} & \colhead{$\chi^2$/d.o.f.}\\
\colhead{ } & \colhead{(1--12 keV)}   & \colhead{}}
\startdata
0.0       &  2.05$\pm$0.02     &  167/21 \\
0.5       &  1.91$\pm$0.03     &   45/22 \\
1.0       &  1.85$\pm$0.04     &   22/21 \\
1.5       &  1.82$\pm$0.04     &   20/20     \\
2.0       &  1.81$\pm$0.05     &   20/19
\enddata
\end{deluxetable}

Many other studies stack X-ray spectra of high-redshift AGNs and yield a
data quality which is sufficient only to derive the average broadband spectral
properties. To compare our data to these studies, we fit a 1PL model only to 
our stacked RBS-QSO spectrum. Moreover, we derive the observed power-law indices 
in the 1--12 keV range depending on different redshifts in 
Table~\ref{table:shift_redshift}. For $z<0.5$, the 1PL fit is a bad representation 
of the data, but once the soft excess is shifted outside the 1--12 keV range this 
model results in a good fit. Above $z\sim 2$ the Compton-hump is expected to be 
redshifted into the 1--12 keV range. Since the Compton hump strengths for each of 
our objects remain unclear, we can only speculate that the added spectrum may show 
further hardening.

%%%%%%%%%%%%%%%%%%%%%%%%%%%%%%%%%%%%%%%%%%%%%%%%%%%%%%%%%%%%%%%%%%%%%%%%%%%%%%%%%%%%%%%%%%%%%%%%%%%%%%%%%%%%%%%%%%%%%%%%%%%%%%%%%%%%%%%%%%

\section{Discussion}
\label{discussion1}
Although we investigate only a small sample, all of our objects show very homogenous
spectral properties: a very smooth soft excess and a hard power-law component. 
Furthermore, in the three high S/N spectra of 
RBS 1055, RBS 320, and RBS 1124 we detect a weak narrow fluorescent Fe K$\alpha$
line.} \cite{page_reeves_2004b} study the spectral properties and the spectral energy
distribution of five high-luminosity radio-quiet QSOs. Their sample spans a redshift 
range of $z=0.24-2.04$. The two objects with a similar redshift range to our objects 
have lower 2--10 keV luminosities than our RBS-QSOs. \cite{piconcelli_jimenez_2005} 
investigate the X-ray properties of a much larger sample of 40 PG quasars in a redshift 
range of 0.036--1.718. The sample contains only five radio-loud QSOs and covers an X-ray 
luminosity of $10^{43}\le L_{\rm 0.5-10}/{\rm[erg\,s^{-1}]} \le 5\times 10^{45}$. Therefore,
PG quasars share very similar properties to our sample and we will compare our results 
mainly to this sample.

None of our studied RBS-QSOs require the presence of a full covering cold absorber
along the line of sight in excess of the Galactic equivalent hydrogen column density. 
This is consistent with the findings of \cite{piconcelli_jimenez_2005} where only 3/35 
radio-quiet PG quasars show evidence for the presence of a cold absorber. The optical 
spectra of our RBS-QSOs reveal AGN-typical features: broad emission lines and UV-excess. 
These type of objects (Seyfert I classification) are, in general, known to agree well 
in optical and X-ray properties with the classical AGN unification scheme 
(\citealt{antonucci_1993}). However, {\em ROSAT's} sensitivity to the soft energy 
band ($<$2.4 keV) selects against X-ray absorbed AGNs. Therefore, our 
sample is biased toward unabsorbed AGNs.

Ionized absorbers are found routinely in $\sim$50\% of Seyfert galaxies 
(\citealt{reynolds_1997}). \cite{piconcelli_jimenez_2005} detect absorption features due 
to ionized gas in 18 out of their 40 PG quasars with similar and lower X-ray
luminosities. They conclude that the detection rate of ionized absorbers appears 
to be independent of X-ray luminosity. For each of our objects we find no evidence for 
the presence of ionized absorbers. The RGS data of the three high S/N RBS-QSOs also 
show no evidence for ionized absorbers. The S/N of our spectra is comparable to 
\cite{piconcelli_jimenez_2005}. Other examples of ionized absorbers in luminous
radio-quiet AGNs are found by \cite{page_reeves_2004b} (one out of five objects 
at $z=0.24$) and \cite{reeves_obrien_2003} ($z=0.18$). The extreme smoothness of 
our RBS-QSO spectra in the soft energy band, with no signs for ionized absorbers, 
is not consistent with these previous studies and may suggest a deficiency of ionized
absorbers in the very high end of the QSO luminosity function.

Partially covering models are only able to sufficiently describe the observed data
of RBS 1124 and RBS 1883. However, in both cases this requires 
$N_{\rm H,cov}\sim 10^{23}$ cm$^{-2}$ which results in very high bolometric luminosity 
(a few 10$^{46}$ erg s$^{-1}$ assuming the bolometric correction to the 2--10 keV 
intrinsic luminosities of \citealt{marconi_risaliti_2004}) and consequently
in super-Eddington ratios of $L_{\rm bol}/L_{\rm edd}$=5.3 for RBS 1124 and 3.9 for
RBS 1883. Therefore, a partially covering model seem to be unlikely to explain 
the X-ray spectra for these two objects.

\subsection{Soft Excess}
The need for a 2PL model to fit the overall X-ray spectra well in all objects 
indicates that all objects have a significant soft excess. Moreover, in all objects 
in the current sample, the soft excess is well fit by a soft power law in CCD detectors. 
The RGS spectra of the three highest S/N spectra are also fit well by a simple power law,
with no strong evidence for narrow emission or absorption lines.
The vast majority of luminous AGNs is known
to show a soft excess (e.g., \citealt{page_reeves_2004b}). For comparison, we derive 
the photon indices by selecting only the energy ranges of 0.5--2 keV and the 2--12 keV,
respectively. We list the results of a single power-law fit to these energy ranges in 
Table~\ref{table:compare}, i.e., these indices are not based on our 1PL or 2PL model fits. 
Our mean soft photon index ($\langle \Gamma_{\rm soft} \rangle=2.36$) is harder than 
the average PG quasar value of $\langle \Gamma_{\rm soft} \rangle=2.73$. Only RBS 320 
and RBS 1883 have soft photon indices comparable to the average PG quasar value, but
these objects also exhibit much softer 2--12 keV photon indices than the other sources.

\cite{miniutti_ponti_2009} evaluate the strength of the soft excess for the PG quasars 
based on a phenomenological blackbody model by computing the contribution of the 
blackbody component to the 0.5--2 keV total flux finding an average of $\sim$30\%.
We list this quantification of the soft excess for our RBS-QSOs in Table~\ref{table:compare}.
Except RBS 1883, all our objects show a much smaller soft excess strength (2\%--9\%) than 
the, on average, lower luminous PG quasars. 

In the PLBB fits of our objects, we derive a mean $\langle k_{\rm B}T_{\rm BB} \rangle=0.12$ keV 
with a dispersion of $\sigma = 0.02$ keV. This is consistent with the values found for AGNs 
with intermediate-mass black holes (and $L_{\rm 0.5-10} \sim 10^{43}$ erg s$^{-1}$) from
\cite{miniutti_ponti_2009} and slightly lower (but still consistent) than the mean value 
for PG quasars ($\langle k_{\rm B}T_{\rm BB} \rangle=0.136$ keV, $\sigma = 0.021$ keV).
As pointed out by several authors (e.g., \citealt{gierlinski_done_2004};
\citealt{ross_fabian_2005}; \citealt{crummy_fabian_2006}), the observation of the relatively 
sharp blackbody temperature distribution of $\sim$0.10--0.15 keV over such wide range of 
X-ray luminosities does not support the idea that the soft excess in AGNs is connected to the 
high-energy tail of the thermal emission from the accretion disk. We find similar blackbody
temperatures in the PLBB fits to our RBS-QSOs. In this context, the soft excess consequently 
has to be explained by a model that produces very similar $T_{\rm BB}$ values independently of 
AGN X-ray luminosity when modeled by a blackbody component, as well as reduces the soft excess 
strength from $\sim$30\% for low and medium X-ray luminous to $\sim$5\% for high X-ray
luminous radio-quiet QSOs.

\begin{deluxetable}{lcccc}
\tabletypesize{\normalsize}
\tablecaption{Additional Fit Results from the RBS-QSO X-ray Spectra\label{table:compare}}
\tablewidth{0pt}
\tablehead{
\colhead{Object} & \colhead{$\Gamma$}   & \colhead{$\Gamma$}   & \colhead{Flux
Ratio}     & \colhead{$L_{\rm X}$} \\
\colhead{Name }  & \colhead{(0.5--2 keV)} & \colhead{(2--12 keV)}  & \colhead{(bb/0.5--2
keV)}  & \colhead{(2--10 keV)}}
\startdata
RBS 1055         & 1.81$\pm$0.02        &  1.62$\pm$0.03       &    0.018$^{+0.002}_{-0.004}$ &  24     
      \\
RBS 320          & 2.69$\pm$0.03        &  2.13$\pm$0.07       &    0.05$\pm$0.01  &  8.7    
       \\
RBS 1124$^{\rm X}$& 2.00$\pm$0.02        &  1.73$\pm$0.03       &    0.05$\pm$0.01  &  8.0   
        \\
RBS 1124$^{\rm S}$& 1.93$\pm$0.03        &  1.69$\pm$0.03       &    0.09$\pm$0.01  &  5.8   
        \\
RBS 1883         & 2.93$\pm$0.20        &  2.36$^{+0.48}_{-0.44}$&    0.26$^{+0.01}_{-0.05}$ &  5.5  
          \\
\hline
Mean             &                      &                      &     &           
       \\
({\em XMM} only)            & 2.36$\pm$0.47        &  1.96$\pm$0.30       &  -- &  11.6
\enddata
\tablecomments{The flux ratio bb/0.5--2 keV quantifies the Galactic absorption-corrected
contribution of the soft excess to the total observed 0.5--2 keV flux; this was computed
using a PLBB fit. The Galactic absorption-corrected 2--10 
                    keV rest-frame luminosity is based on the best-fit model of each object 
                    (in units of $10^{44}$ erg s$^{-1}$). Superscript ``X'' refers to 
                    the {\em XMM-Newton} observation of RBS 1124, while ``S'' indicates  
                    the {\em Suzaku} observation. For the error of the mean photon indices 
                    we list the dispersion $\sigma$.}
\end{deluxetable}

\subsection{Hard X-ray Continuum}
The mean 2--12 keV photon index of our RBS-QSOs is $\langle \Gamma_{\rm hard}
\rangle=1.96$ ($\sigma = 0.30$). Taking into account that this value is only 
based on four objects, it agrees well with the mean hard power-law component 
of radio-quiet PG quasars ($\langle \Gamma_{\rm hard} \rangle=1.89$, 
$\sigma \sim 0.36$) and the QSO sample of \citealt{page_reeves_2004b}
($\langle \Gamma_{\rm hard} \rangle=1.93\pm0.02$). \cite{piconcelli_jimenez_2005}
find some significant ($>$99\% confidence level) correlations among the 2--12 keV 
power law/luminosity and QSO physical properties. Our objects, testing these 
correlations at the high end of the QSO luminosity function, fit well into the
$M_{\rm BH}$--$L_{\rm 2-10}$ and FWHM(H$\beta$)--$\Gamma$ correlation. They are
somewhat consistent with the $\Gamma$--$M_{\rm BH}$ correlation, but do not fit 
well into the FWHM(H$\beta$)--$L_{\rm 2-10}$ correlation
(higher FWHM(H$\beta$) values are expected for our luminosities).
The hard photon indices of our individual objects are consistent with the 
$\Gamma$--$L_{\rm bol}/L_{\rm edd}$ correlation (e.g., \citealt{lu_yu_1999}; 
\citealt{shemmer_brandt_2008}).

\subsection{Origin of the Fe Line}
\label{origin_of_feline}
Three out of our four studied objects have spectra with a sufficiently high S/N
to detect an Fe line whose energies are consistent with a neutral
fluorescent Fe K$\alpha$ line at 6.4 keV. A line energy of 6.7 keV from He-like
Fe emission can be excluded in the individual and stacked {\em XMM-Newton}
observations with a statistical significance of 1.5$\sigma$ to $>$3$\sigma$.
All of the lines are either barely resolved or nearly resolved 
($\sigma_{\rm Fe, rest}=0.15-0.5$ keV). The rest-frame equivalent widths are  
70, 130, and 140 eV ({\em XMM-Newton} observations only) and are consistent between 
the individual objects taking into account the uncertainties. The average equivalent 
width of the individual observations of $\langle$EW$\rangle =110$ eV is in very
good agreement with the derived equivalent width of the stacked RBS-QSO spectrum
(EW$ = 120\pm 60$ eV).

\cite{jimenez-bailon_piconelli_2005} investigate the properties of the fluorescent Fe
K$\alpha$ line in the PG quasar sample of \cite{piconcelli_jimenez_2005}. A narrow
Fe line is found in 50\% of all radio-quiet PG quasars. The vast majority of their
narrow lines have an energy centroid consistent with $E=6.4$ keV, indicating matter 
in low ionization states (\ion{Fe}{1}--\ion{Fe}{17}) which is expected to be located at large
distances from the central X-ray source. However, at luminosities $L_{\rm 2-10}$ 
greater than $3 \times 10^{44}$ erg s$^{-1}$, they only have one radio-quiet 
object with sufficient S/N to detect a line.  Our study adds three data points 
in this poorly studied luminosity range. The detection of only narrow, yet barely 
resolved, Fe lines from neutral/low ionization stages agree well with the results 
of PG quasars and other QSO samples (e.g., \citealt{page_reeves_2004b}; 
\citealt{porquet_reeves_2004}). Therefore, narrow Fe emission from low ionization 
states is a common spectral feature for low to very luminous AGNs and suggests that 
a common mechanism or set of mechanisms governing transmission and/or reflection
processes must exist across a wide range of luminosities.

We compute the FWHM velocity line width for the Fe K$\alpha$ lines in each of our
objects based on the 2PLG fits (Table~\ref{powerlaw}). For RBS 1124, we used the 
line width from the joint {\em Suzaku} and {\em XMM-Newton} fit since it gives
slightly better constraints and the values from both observations agree within their
uncertainties.
Assuming that the material in which the Fe line originates is in a Keplerian orbit,
the FWHM velocities of $53,000\pm 20,000$ km s$^{-1}$ (RBS 1055) and 
16,000$^{+11,000}_{-10,000}$ km s$^{-1}$ (RBS 1124) correspond to distances of only 
2.4$^{+3.8}_{-1.1}$  and 4.1$^{+24.9}_{-2.7}$ light days, respectively, from the central 
supermassive black hole (using $\langle v^2 \rangle = \frac{3}{4}v^2_{\rm FWHM}$,
\citealt{netzer_1990}). These distances are even closer to the black hole than the
Broad Line Region (BLR) that is believed to emit the broad optical emission lines. 
If the velocities are Keplerian the \ion{Mg}{2} emission in RBS 1055 and H$\beta$ 
in RBS 1124 originate from radii of $\sim$220 and $\sim$58 light days, respectively. 
For RBS 320 the lower 90\% confidence limit is $r>1.2$ light days
($<$75,000 km s$^{-1}$, with H$\beta$ emission originating at a distance of
$\sim$1300 light days). 

It is somewhat surprising to detect neutral Fe emission from a radius of only a few 
light days, hence closer than the BLR, in the most luminous QSOs. It is reasonable 
to expect that the material may be highly ionized at these radii. Shielding processes 
could sustain neutral material so close to the central engine. However, a partial
covering model does not fit the data of RBS 1055 well and is also unlikely for RBS
1124 (see discussion in Section~\ref{discussion1}). Another possible scenario of 
maintaining neutral material very close to very luminous AGNs may be non-isotropic 
emission of the ionizing continuum radiation.

The PG quasar sample shows that the detection of relativistically broadened Fe lines 
is not common for high luminous AGNs except in three cases. Our non-detections are 
consistent with the notion that relativistically broadened Fe lines are rare
at the high end of the QSO luminosity function. This finding seems to also apply 
to high luminosity type II QSOs for which \cite{krumpe_lamer_2008} do not 
detect a relativistically broadened Fe line (EW upper limit of 1 keV) in the 
stacked spectrum of type II QSOs, while lower luminosity AGNs do show evidence for 
such a line (EW$ = 2.0^{+0.6}_{-0.7}$). Possible selection effects due to the 
limited S/N in the spectra of the luminous QSOs may hamper the detection of broad 
Fe lines. We determine the 90\% confidence level upper limit of a {\tt laor} 
component for each of RBS 1055, RBS 320, and RBS 1124 and find that broad Fe lines 
with EWs up to $\sim 140-400$ keV could be hidden in the spectra. The three 
detected broad Fe lines in the PG quasar sample have EW$ = 100-500$ eV (in objects with 
$L_{\rm 2-10} \lesssim 3 \times 10^{44}$ erg s$^{-1}$). De la Calle P{\'e}rez et 
al.~(2010)\nocite{calleperez_longinotti_2010} investigate a sample of radio-quiet 
AGNs and QSOs and detect only for 2 QSO ($<$6\% of all QSOs) a relativistically 
broadened Fe line. They also conclude that the measured EW is always below 300 eV 
and has a mean value of $\langle$EW$\rangle = 143 \pm 27$ eV. An extreme case is 
the radio-quiet very luminous QSO PG 1247+267 (\citealt{page_reeves_2004b}, 
$z=2.04$, $L_{\rm 2-10} = 9 \times 10^{45}$ erg s$^{-1}$) which shows an apparently 
broad line with EW$ = 421\pm 215$ eV. Therefore, the expected EWs of broad Fe lines 
based on previously published results are not detectable in our 
RBS-QSO spectra. At low redshifts, low luminosity Seyfert AGNs show a 
relativistically broadened Fe line in $\sim$50\% of the objects 
(\citealt{nandra_oneill_2007}). However, the major fraction of the detected lines 
in these high S/N spectra has EW$ < 200$ eV and therefore falls below our detection 
limit for a broad Fe line. Consequently, our results do not allow a conclusion 
on similar or different physical properties in the accretion disk very close to the 
central engine in low and high luminous AGNs.

\subsection{Comparison to Low-$z$ AGNs}
Low-$z$ AGNs vary significantly in their individual X-ray spectral components
and feature a wide range of frequency of partial coverers, warm absorbers, soft
excess shapes and strengths, and/or Compton reflection hump strengths. It is 
therefore not straightforward to compare our results to a somewhat typical 
low-$z$ AGN to evaluate what changes have to be applied to the X-ray spectrum 
(except the much higher luminosity) to turn a low-$z$ AGN into one of the most
luminous QSOs.

Since we are left with the situation where our RBS-QSO spectra can be well fitted with
power law and self-consistent reflection disk models (EPIC and RGS data), we decided
to compare our findings with \cite{crummy_fabian_2006}. They used a sample of 34 
mainly low-redshift Seyfert I galaxies and type I AGNs to show that in 25 out of 
34 sources, relativistically blurred disk reflection models are a significantly
better fit than a PLBB model. All three of our high S/N RBS-QSO observations are
better fitted with reflection disk models than PLBB models. However, a 2PL fit is 
almost as good as a reflection model, because the extremely smooth soft excess
in the EPIC and RGS data is well fitted by a soft power law. A reflection model has
to accomplish the modeling of the smooth soft excess similar to a power law and 
requires a high degree of relativistically blurring to smear out the sharp soft 
X-ray emission lines caused by reflection from the disk. This is achieved by either 
very high ionization parameters, small inner disk radii, or high emissivity indices
(causing the emission of the major fraction of the radiation within very small inner
radii), or a combination of those. In this case, the limited parameter space no longer 
allows modeling of the Fe line in a satisfying way as in REFL1 and REFL2, requiring 
us to include the extra, narrow Fe component as in REFL1G and REFL2G.
Moreover, we find a degeneracy in the disk reflection model between accretion disk
states at seemingly opposite physical extremes: REFL1 has a very low ionization
parameter and accounts only for $\sim$5\%--20\% of the total 0.2--12 keV flux, while 
REFL2 exhibits a very high disk ionization parameter ($\sim$35\%--75\% of the total 
output).

Some objects in the \cite{crummy_fabian_2006} sample (e.g., TON S180, where the soft
excess is also better modeled with a power law rather than a blackbody model) also 
have a very smooth soft excess and require high ionization parameters and high 
emissivity indices. \cite{crummy_fabian_2006} allow the emissivity index to vary 
freely for all their sources and find an index higher than 6 for 22 sources. Only 
two of their sources are consistent with an index below or equal to 3. Leaving the 
emissivity index free to vary in our fits results in unconstrained values in most of 
the cases due to the limited S/N. Only for RBS 320 and REFL2 of the RBS 1124 {\em 
Suzaku} observation we achieve well constrained emissivity indices higher than 6.   

\cite{crummy_fabian_2006} find very small inner disk radii indicating fast rotating
Kerr black holes which is in agreement with our data. Interestingly, their 0.3--12 keV 
flux fraction for the disk reflection models is always above 0.25 which contradicts 
our models REFL1 and REFL1G (low ionization parameter; flux fraction $\sim$5\%--20\%). 
The best-fit ionization parameters in \cite{crummy_fabian_2006} cover a wide range 
including the values found for our RBS-QSOs. They also find that most sources have 
similar spectral indices with both models (PLBB and disk reflection), while several 
sources have higher indices with the disk reflection model. The ranges of our 
PLBB and REFL1G photon indices agree well with the correlation given by 
\cite{crummy_fabian_2006}. For REFL2G we find lower spectral indices than with
a PLBB. 

None of the degenerate disk reflection scenarios, i.e., low/high ionization parameter, 
for our RBS-QSOs can be favored or ruled out. The limited S/N of the RGS data also 
does not allow any constraints. If the soft excess in these objects is not caused by 
reflection in the innermost disk, there is in general no need for relativistical disk 
reflection models in our high luminosity QSOs as the observed Fe line does not originate 
from the inner disk although we cannot rule out the possibility that there may exist 
a weak broad Fe line. Either the basic assumptions of reflection models (e.g., 
geometrically flat accretion disk, gas of constant density, etc.) are not valid for the 
most luminous QSOs, or different sets of models are required to describe their physics. 
If, on the other hand, reflection models turn out to be the right interpretation for very 
luminous QSOs, this would support the assumption by \citealt{page_reeves_2004b} that high 
luminosity radio-quiet QSOs have identical X-ray continuum properties as those found in 
lower-luminosity AGNs and Seyfert I galaxies and therefore similar physical processes are 
likely to govern accretion disks over a wide range of luminosities and accretion rates.

\subsection{Comparison to High-$z$ QSOs}
In this section we want to explore if our stacked high S/N RBS-QSO spectrum can 
serve as a template spectrum for low S/N spectra of luminous QSOs at higher 
redshifts ($z \gtrsim 1$). The PG quasar sample of \cite{piconcelli_jimenez_2005} 
contains five objects with $z>1$ (up to $z=1.7$) and X-ray luminosities of 
$L_{\rm 2-10} \sim (15-130) \times 10^{44}$ erg s$^{-1}$. These objects have an 
average 2--12 keV photon index $\langle \Gamma_{\rm 2-12}\rangle = 2.1$ with a 
typical error of $\pm0.3$. Our photon index for the stacked RBS-QSO spectrum shifted 
to $z=1.0$ is $\Gamma_{\rm 1-12}=1.85\pm 0.04$. Taking into account the slightly 
different energy range for the determination of the photon indices and the 
involved uncertainties, both values are in good agreement. 

Although we are aware of the large uncertainties in using a sample of only three 
objects, we derive the average soft and hard photon indices for the 
three $z>1$ objects of \cite{piconcelli_jimenez_2005}. Their 
$\langle \Gamma_{\rm soft}\rangle = 3.4$ and $\langle \Gamma_{\rm hard}\rangle = 1.4$ 
agree well with the 2PL fit to our stacked RBS-QSO spectrum 
($\Gamma_{\rm soft} = 3.4^{+0.8}_{-0.6}$, $\Gamma_{\rm hard} = 1.7^{+0.1}_{-0.2}$).

The limited S/N for luminous QSOs at high redshifts restricts the knowledge that 
can be gained from these objects. By stacking the individual X-ray spectra the 
S/N is improved significantly and the added spectrum can be used for further analysis 
which is not feasible on individual spectra. However, the interpretation of stacked 
spectra \textit{a priori} assumes that the intrinsic X-ray properties in the sample 
are very similar. Radio-quiet AGNs correspond to the vast majority ($\sim$90\%) of 
all AGNs. Although limited to only four objects at the present, our RBS-QSOs are 
found to have surprisingly homogenous properties in their X-ray spectra. This confirms 
the stacking technique assumption of very similar properties in the individual spectra 
at least for very luminous QSOs. 

In one of the largest samples, \cite{mateos_carrera_2010} stack the spectra of 487 
type I AGNs in the {\em XMM-Newton} Wide Angle Survey (XWAS) in different redshift 
bins. The two highest redshift bin subsamples with $\langle z \rangle = 2.01, 2.56$ 
and $L_{\rm 2-10} \sim 4 \times 10^{44}$ erg s$^{-1}$ both yield a best-fit photon 
index of $\langle \Gamma \rangle = 1.81\pm 0.05$. Our stacked spectrum redshifted to 
$z=2$ is best described by a single power-law fit with $\Gamma = 1.81\pm 0.05$. 
Although both results are in perfect agreement, \cite{mateos_carrera_2010} mention that 
their value could be biased in such a way that at high redshifts the efficiency of 
detecting hard objects is increased, while for soft sources it decreases. They also 
point out that above $\sim$ 10$^{44}$ erg s$^{-1}$, the average photon index seems to 
remain fairly constant over two orders of magnitude in luminosity. 
\cite{mateos_carrera_2010} find little or no evidence for X-ray absorption (only 
$\sim$3\% of the sources show significant full covering cold absorber along the line 
of sight) and a soft excess model (PLBB model versus single power law) represents 
the data better in $\sim$36\% of the objects below $z=0.5$ ($F$-test significance 
$\ge$99\%, note that their median number of total EPIC counts in the 0.2--12 keV band 
in the individual spectra is $\sim$300 counts). Our objects fit well the correlation 
of the soft excess luminosity versus intrinsic power-law continuum luminosity 
presented by \cite{mateos_carrera_2010}.  

\cite{corral_page_2008} stack the X-ray spectra of 606 type I AGNs in the XWAS and 
study the properties of the average Fe line. For the highest luminosity sub-sample 
$L_{\rm 0.5-2} \sim (6-660) \times 10^{44}$ erg s$^{-1}$, they find an average Fe line 
EW of 50$^{+50}_{-40}$ eV. Our average Fe line EW from the stacked spectrum is 
EW$ = 120\pm 60$ eV, which is consistent with their value within the uncertainties. 
\cite{chaudhary_brusa_2010} use 507 AGNs from the 2XMM catalog and find for subsamples 
with $L_{\rm 2-10} \gtrsim 5 \times 10^{44}$ erg s$^{-1}$ in a redshift range of $z=0.8-5.0$ 
upper limits on the EW also consistent with our values. These new stacking results and 
our direct measurements for the RBS-QSOs significantly disagree with the large EWs found 
by \cite{streblyanska_hasinger_2005} and \cite{brusa_gilli_2005} using the stacking 
technique.

\subsection{X-ray Baldwin Effect}
A unique and important contribution of our study is that we can test the X-ray 
Baldwin effect, the anti-correlation between X-ray luminosity and EW of the narrow 
Fe line (\citealt{iwasawa_taniguchi_1993}), for the most luminous QSOs. Several 
previous studies have lively debated the existence and the significance of the 
X-ray Baldwin effect. For example, \cite{nandra_george_1997}, \cite{page_obrien_2004a}, 
and \cite{bianchi_guainazzi_2007} confirm the presence of the X-ray Baldwin effect 
in different samples of AGNs. \cite{jimenez-bailon_piconelli_2005} question the 
significance of the effect by finding only a 30\% probability that the EW and the 
2--10 keV luminosity are correlated. They and \cite{jiang_wang_2006} both 
argue that the effect is mainly due to the radio-loud sources in the sample. 

\begin{figure}[t]
  \centering
 \resizebox{\hsize}{!}{
 \includegraphics[bbllx=69,bblly=372,bburx=534,bbury=700]{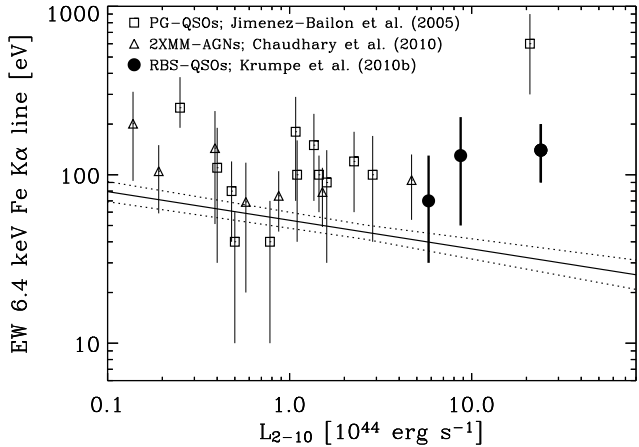}}
  \caption{X-ray Baldwin effect: neutral Fe line equivalent width (EW) against the 2--10 keV 
luminosity. The solid line shows the best-fit to the AGN sample from Bianchi et al. (2007); 
the dotted lines represent the error on their fitted correlation. 
Filled circles show our RBS objects.}
         \label{baldwin}
\end{figure}

The main problem is that the fitted correlation for the X-ray Baldwin effect 
comes mainly from AGNs with 
$10^{42}\le L_{\rm 2-10}/{\rm[erg\,s^{-1}]} \le 4\times 10^{44}$. 
\cite{jimenez-bailon_piconelli_2005} and \cite{bianchi_guainazzi_2007} note that 
more luminous radio-quiet QSOs with high S/N are needed for a solid analysis of 
the X-ray Baldwin effect. Our study adds three data points to this very poorly 
studied luminosity range, with EW$ = 70-140$ eV ($\langle$EW$\rangle =110$). This 
is still consistent with the PG quasar average value of 
$\langle$EW$\rangle =80 \pm 20$ eV with an dispersion of $\sigma_{\rm EW} < 40$ eV
(\citealt{jimenez-bailon_piconelli_2005}). Based on our 2--10 keV luminosities, 
we calculate the expected EW from the best-fit of \cite{bianchi_guainazzi_2007} 
(their Equation~1, sample of radio-quiet AGNs). The expected values of 
EW$ \sim (31-40)\pm 6$ eV are significantly lower than what we measure in our 
RBS-QSOs (Figure~\ref{baldwin}). Except for RBS 1124 ({\em XMM-Newton} and {\em Suzaku} observations), 
the values disagree at the $>$90\% confidence level. However, other studies also 
find much higher EWs (as predicted by the X-ray Baldwin effect) of $\sim$100 eV for 
$L_{\rm 2-10} \gtrsim 10^{44}$ erg s$^{-1}$ (e.g., \citealt{chaudhary_brusa_2010}; 
\citealt{reeves_turner_2000}; \citealt{page_schartel_2004c}). Figure~\ref{baldwin}
only contains objects from \cite{jimenez-bailon_piconelli_2005} that have a line 
energy consistent with $E=6.4$ keV. Their sample and the sample of 
\cite{bianchi_guainazzi_2007} have 25 PG quasars in common. Although consistent 
within their uncertainties, the 
derived EW values for detected Fe lines by \cite{bianchi_guainazzi_2007} are systematically lower because, in contrast 
to \cite{jimenez-bailon_piconelli_2005}, they fixed the line energy to $E=6.4$ keV 
(S. Bianchi, private communication).

Based on our study of RBS-QSOs, and on the $L_{\rm X}-$EW relations of 
\cite{chaudhary_brusa_2010} and \cite{jimenez-bailon_piconelli_2005}, we conclude 
that the X-ray Baldwin effect, the proposed anti-correlation between X-ray luminosity and EW 
of the narrow Fe line, is only present up to $L_{\rm 2-10} \sim 10^{44}$ erg s$^{-1}$. 
Above this luminosity threshold, there is convincing evidence for a flattening of the 
X-ray Baldwin effect or even a luminosity independent value of the equivalent width 
of $\langle$EW$\rangle\sim 100$ eV. We hereby redefine the X-ray Baldwin effect not 
only as the proposed anti-correlation between $L_{\rm X}$ and EW but also as the flattening 
above $L_{\rm 2-10} \sim 10^{44}$ erg s$^{-1}$ to cover a luminosity range as wide as 
possible.  With current data samples, we are not 
able to tell if this flattening applies only to radio-quiet QSOs or for all QSOs. 
However, the flattening would explain the discrepancy in the significance of the X-ray 
Baldwin effect when different luminosity ranges are explored, such as the very high 
luminosity range for PG quasars in \cite{jimenez-bailon_piconelli_2005}.

The most fundamental question concerning the X-ray Baldwin effect is the physical 
cause why the behavior changes above $L_{\rm 2-10} \sim 10^{44}$ erg s$^{-1}$.
Depending on the origin of the Fe line, different explanations are suggested. 
If the flattening or constant EWs above $L_{\rm 2-10} \sim 10^{44}$ erg s$^{-1}$ prove 
to be solid, either some physical process reaches its limit at this luminosity or 
two different mechanisms/origins of the Fe line are causing the X-ray Baldwin effect. 
For our objects, the X-ray Baldwin effect is most likely not caused by the notion 
that an increase in X-ray luminosity progressively further ionizes the upper layers 
of the disk and therefore decreases the strength of the line 
(\citealt{ross_fabian_2005}) because the main contribution to the observed Fe line 
does not seem to be located in the inner disk region. Other possible explanations
are that with increasing luminosity, the inner radius of the torus surrounding the 
supermassive black hole increases up to a certain limit or the covering factor or 
thickness of the torus is decreasing (\citealt{page_obrien_2004a}). If the origin 
can be constrained to be in the torus, this luminosity-dependent covering factor 
may be directly linked to the decrease of the fraction of obscured AGNs (\citealt{ueda_akiyama_2003}).
Both scenarios 
lead to a lower amount of reprocessed radiation and therefore lower observed neutral 
reflection and narrow Fe emission line strengths. However, the model still has to 
explain the high accretion rates found in RBS-QSO that seem to be sustained over 
a long period of time.

%%%%%%%%%%%%%%%%%%%%%%%%%%%%%%%%%%%%%%%%%%%%%%%%%%%%%%%%%%%%%%%%%%%%%%%%%%%%%%%%%%%%%%%%%%%%%%%%%%%%%%%%%%%%%%%%%%%%%%%%%%%%%%%%%%%%%%%%%%
\section{Conclusions}

We select the 12 most luminous radio-quiet QSOs from the {\em ROSAT} 
Bright Survey (RBS) and present the first mini-survey of four of these 
sources. They each have $L_{\rm 0.5-2} \ge 10^{45}$ erg s$^{-1}$, 
radio-to-optical flux densities $R<10$, and allow us to study the 
accretion disk properties at the very high end of the QSO luminosity 
function. These four objects have $z=0.21-0.55$ and each have been 
observed with {\em XMM-Newton}; RBS 1124 has also been observed with 
{\em Suzaku} (\citealt{miniutti_piconcelli_2010}). Using emission lines 
from optical spectra, we estimate black hole masses of 
log($M_{\rm BH}/M_{\odot}$)$=8.06-8.88$ and Eddington ratios of 
$L_{\rm bol}/L_{\rm edd}=0.10-1.43$ for these objects. Therefore, the outstanding 
luminosities are a combination of high but not extreme black hole masses 
and Eddington ratios. The {\em XMM-Newton} 0.5--2 keV fluxes vary only 
$\sim$5\%--20\% from the {\em ROSAT} fluxes, while the {\em Suzaku} and 
{\em XMM-Newton} observations of RBS 1124 show a 50\% flux increase in 
only 13 months. The optical-to-X-ray spectral indices of our objects
cover the same range found in other {\em ROSAT} selected AGN samples.

All 0.2--12 keV {\em XMM-Newton} spectra show very homogenous properties, 
implying that the physical properties in the individual accretion disks 
are most likely very similar in this uniformly selected sample. The data 
of the three high S/N objects are well described by either (1) a model 
with two power laws or (2) an incident power law and its reflection off 
the accretion disk very close to the black hole, but in all cases, we 
also have to model the presence of a narrow/barely resolved Fe line 
originating in neutral material light-days or further from the central 
X-ray source with an additional Gaussian line profile.

Reflection models alone are not able to explain the strength of the 
Fe line. Two degenerate disk reflection model scenarios, which lie in 
seemingly very different areas of parameter space, are found for the 
high S/N spectra: (1) a disk with a very low ionization parameter (REFL1) 
with a reflection component which makes up only $\sim$5\%--20\% of the total 
output or (2) a disk with a very high ionization parameter 
(REFL2--as almost expected for very luminous QSOs) with a reflection component 
that accounts for $\sim$35\%--75\% of the total output. Due to the weak contribution 
of the reflection component in REFL1, even the limited S/N RGS data do not allow 
us to favor either model. This degeneracy in the disk reflection model has to be 
broken by either more advanced models (maybe disk reflection models are even
inappropriate for very luminosity radio-quiet QSOs) or extremely high-resolution 
spectra such as those which will be obtained by upcoming X-ray missions such as 
{\em Astro-H} (\citealt{takahashi_kelley_2008}) and {\em IXO} 
(\citealt{white_parmar_2009}) to identify the physical properties of the 
accretion disk in the most luminous radio-quiet QSOs. Both scenarios are 
consistent with the presence of fast rotating Kerr black holes.

The soft excess in each source is well fitted by a power law, and its strength 
(2\%--26\% of the total 0.5--2 keV emission) is weaker than in lower luminous 
radio-quiet AGN/QSOs where the soft excess is typically 30\% of the total 
0.5--2 keV emission. No significant evidence for neutral absorption in excess 
of the Galactic absorption and for partial/full-covering absorption by ionized 
material is found. The RGS data confirm the extreme smoothness of the soft excess 
and the absence of ionized absorbers.  

The major fraction of the Fe line emission in our three high S/N RBS-QSOs has to 
come from lowly-ionized or neutral material located only a few light days (or further) 
out, but a superimposed very weak relativistically broadened Fe line from the inner disk 
cannot be ruled out (90\% confidence level upper limits from fitting {\tt laor} components: 
$\sim$0.14--0.40 keV). It remains unclear how a considerable fraction of the line-emitting
material can avoid becoming highly ionized so close to the central X-ray source of the 
most luminous QSOs.   

We also test the X-ray Baldwin effect in the poorly studied high luminosity range.
The detected Fe line equivalent widths of 70--140 eV from our three high S/N RBS-QSO 
spectra are significant higher than expected by extension of the X-ray Baldwin effect 
from lower luminosity regimes (\citealt{bianchi_guainazzi_2007}). Based on our results 
and the findings by \cite{jimenez-bailon_piconelli_2005} and \cite{chaudhary_brusa_2010}, 
we claim that the X-ray Baldwin anti-correlation is only present in radio-quiet AGNs with 
$L_{\rm 2-10} \lesssim 10^{44}$ erg s$^{-1}$ and above this luminosity, the observed EW 
flattens ($\langle$EW$\rangle \sim 100$ eV). 

Although our present sample is only limited to four objects, we can confirm the stacking 
technique assumption of having very similar intrinsic properties in the individual spectra 
for very luminous QSOs. The most luminous radio-quiet RBS-QSO are very likely to be the 
standard cases of luminous radio-quiet QSO at high redshifts. They can serve as a medium-$z$ 
reference for the interpretation of the low S/N spectra of QSOs with similar luminosities 
at higher redshifts, which are routinely found by {\em XMM-Newton} and {\em Chandra} surveys. 
Knowledge about intrinsic spectral components and the scatter in their parameters values 
is important e.g., when estimating required observation times of similar objects
using current or future missions (the reader is referred to the best-fit results for a 
model with two power laws plus a Gaussian Fe line profile in Section~\ref{stacked_spectrum}).

\acknowledgments
Mirko Krumpe is supported by the NASA grant NNX08AX50G. Amalia Corral acknowledges 
financial support from ASI (grant n.I/088/06/0). We especially thank 
Richard E.\ Rothschild and Takamitsu Miyaji for helpful discussions and Peter Curran and Mathew Page
for offering their help with the reduction of the {\em Swift} XRT spectrum for 
RBS 320.

This work is based on observations obtained with {\em XMM-Newton}, 
an ESA science mission with instruments and contributions directly funded by 
ESA Member States and NASA. The {\em ROSAT} Project was supported by the 
Bundesministerium f{\"u}r Bildung und Forschung (BMBF/DLR) and the Max-Planck-Gesellschaft (MPG).
This work made use of data supplied by the UK Swift Science Data Centre at the
University of Leicester and of data obtained from the 
{\em Suzaku} satellite, a collaborative mission between the space agencies of Japan
(JAXA) and USA (NASA).


%% in parentheses, after the keyword, but they will not be verified.

%{\it Facilities:} \facility{ROSAT}, \facility{XMM-Newton}, \facility{Suzaku} , \facility{Swift}.

%%%%%%%%%%%%%%%%%%%%%%%%%%%%%%%%%%%%%%%%%%%%%%%%%%%%%%%%%%%%%%%%%%%%%%%%%%%%%%%%%%%%%%%%%%%%%%%%%%%%%%%%%%%%%%%%%%%%%%%%%%%%%%%%%%%%%%%%%%

\end{document}